\documentclass[10pt,journal,compsoc]{IEEEtran}
\usepackage{amsmath,amsfonts}
\usepackage{algorithmic}
\usepackage{algorithm}
\usepackage{array}
\usepackage[caption=false,font=normalsize,labelfont=sf,textfont=sf]{subfig}
\usepackage{textcomp}
\usepackage{stfloats}
\usepackage{url}
\usepackage{multicol}
\usepackage{verbatim}
\usepackage{graphicx}
\usepackage{cite}
\usepackage{amsmath} 
\usepackage{array} 
\usepackage{makecell} 
\usepackage[utf8]{inputenc}
\usepackage{hyperref}
\usepackage{amsfonts}
\usepackage{enumitem}
\usepackage{multirow}

\newcolumntype{?}[1]{!{\vrule width #1}}
\newdimen\uulinesep
\begin{document}

\title{Agent Centric Operating System -- a Comprehensive Review and Outlook for Operating System}

\author{
    \IEEEauthorblockN{
    Shian Jia\IEEEauthorrefmark{2}, 
    Xinbo Wang\IEEEauthorrefmark{2}, 
    Mingli Song 
    and Gang Chen 
    }
    \\
\thanks{\IEEEauthorrefmark{2} Equal Contribution.}
\IEEEauthorblockA{ College of Computer Science and Technology \\ Zhejiang University, Hangzhou 310058 \\  \href{mailto:csjsa@zju.edu.cn, xinbowang@zju.edu.cn, brooksong@zju.edu.cn, cg@zju.edu.cn}{\{csjsa, xinbowang, brooksong, cg\}@zju.edu.cn}}
}



\IEEEtitleabstractindextext{%
\begin{abstract}
The operating system (OS) is the backbone of modern computing, providing essential services and managing resources for computer hardware and software. This review paper offers an in-depth analysis of operating systems' evolution, current state, and prospects. We begin with an overview of the concept and significance of operating systems in the digital era. In the second section, we delve into the existing released operating systems, examining their architectures, functionalities, and the ecosystems they support. We then explore recent advances in OS evolution, highlighting innovations in real-time processing, distributed computing, and security. The third section focuses on the new era of operating systems, discussing emerging trends like the Internet of Things (IoT), cloud computing, and artificial intelligence (AI) integration. We also consider the challenges and opportunities presented by these developments. This review concludes with a synthesis of the current landscape and a forward-looking discussion on the future trajectories of operating systems, including open issues and areas ripe for further research and innovation. Finally, we put forward a new OS architecture.
\end{abstract}

}

\IEEEdisplaynontitleabstractindextext

\maketitle

\section{Introduction}
\label{sec:Intro}
\IEEEPARstart{T}{he} OS stands as the cornerstone of contemporary computing, serving as the intermediary between hardware and software, and providing the essential services required for effective resource management and application execution. This paper provides an in-depth examination of the evolutionary trajectory, present status, and prospective developments of operating systems. It begins with exploring the fundamental concept and importance of operating systems in the digital age, emphasizing their pivotal role in facilitating technological advancements.

Subsequently, we delve into a detailed analysis of existing operating systems, covering a broad spectrum from embedded systems to desktop, mobile, and cloud-based platforms. Each type of operating system has evolved to cater to specific requirements and operational contexts, demonstrating remarkable adaptability and innovation. Embedded systems are optimized for real-time processing and resource-constrained environments, making them essential in areas such as smart home devices, automotive electronics, and industrial automation. Desktop operating systems provide robust functionalities and user-friendly interfaces for various applications. Mobile operating systems focus on power efficiency and user experience, supporting on-the-go connectivity and applications. Server operating systems leverage extensive resources to deliver scalable, resilient services for enterprise and data-intensive tasks. This comprehensive overview underscores the specialized nature of operating systems across different computing domains.

\begin{figure}[!t]
\centering
\includegraphics[draft=false, width=\columnwidth]{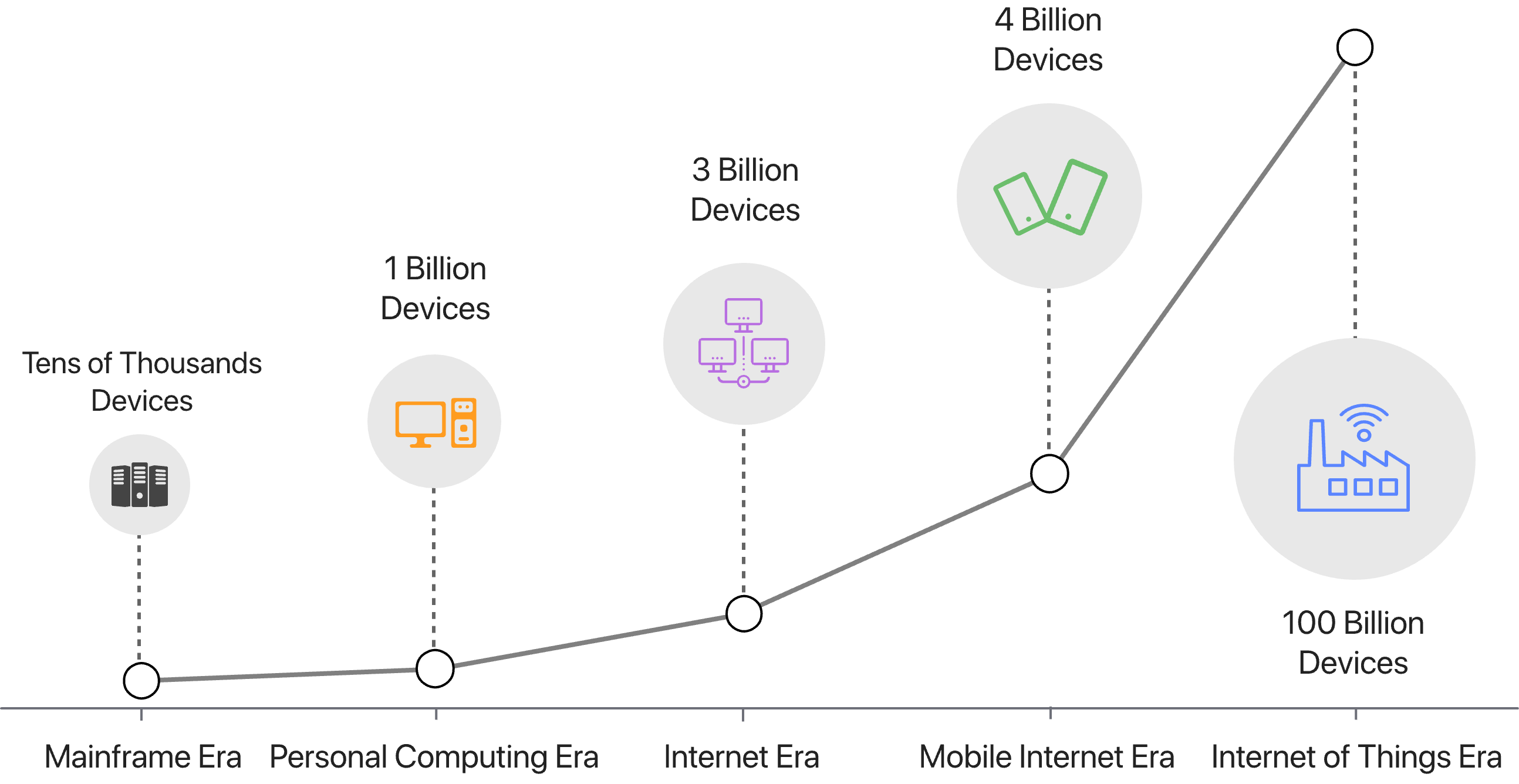}
\caption{\textbf{Evolution of Computing Eras and Corresponding Growth in Connected Devices.} This diagram illustrates the progression through various computing eras, highlighting the exponential increase in connected devices from tens of thousands during the Mainframe Era to an estimated 100 billion in the Internet of Things era. Each era represents significant technological advancements leading to greater connectivity and device proliferation.}
\label{Evolution of Computing Eras and Corresponding Growth in Connected Devices.}
\end{figure}

In the new era of operating systems, we introduce Agent-Centric Operating System (ACOS), a revolutionary concept that embodies the principles of modularity, adaptability, and cross-platform compatibility. ACOS proposes the abstraction of all system components into agents, fostering a flexible and scalable architecture capable of adapting to the unique characteristics of various resource platforms. This approach simplifies system maintenance and updates and enhances the efficiency and responsiveness of operating systems across a wide array of devices.

In this paper, we aim to provide a comprehensive overview of the current state and future directions of OSs. Our contributions are multifaceted. First, we thoroughly analyze existing operating systems, encompassing embedded, desktop, mobile, and server platforms, highlighting their unique features and applications. Second, we explore recent advancements in critical areas such as AI-optimized scheduling, heterogeneous computing, and security and discuss their implications for the evolution of operating systems. Third, we introduce ACOS, a novel operating system concept that emphasizes modularity, adaptability, and cross-platform compatibility, and outline its potential to transform the landscape of operating system design. Finally, we identify current challenges in the field and propose future research directions to address these issues, ensuring that operating systems continue to meet the evolving needs of users and technology.

The organization of this paper starts with an overview of operating systems' significance and foundational concepts in section~\ref{sec:Intro}. Following this in section~\ref{sec:Existing OS}, we offer a detailed review of existing operating systems and their key features. Subsequently, in section~\ref{sec: Recent Advances} we discuss recent advancements and emerging trends in the field, underscoring the transformative impact of technological innovations. Then presented in section~\ref{New Era of OS} is the introduction of ACOS and its innovative design philosophy, followed by a critical analysis of the challenges and opportunities associated with developing next-generation operating systems. Finally, in section~\ref{sec: Conclusion and Outlook} we conclude the paper by summarizing the current landscape of operating system development and suggesting avenues for future research and innovation.

\begin{figure*}[!t]
\centering
\includegraphics[width=\textwidth, keepaspectratio, draft=false]{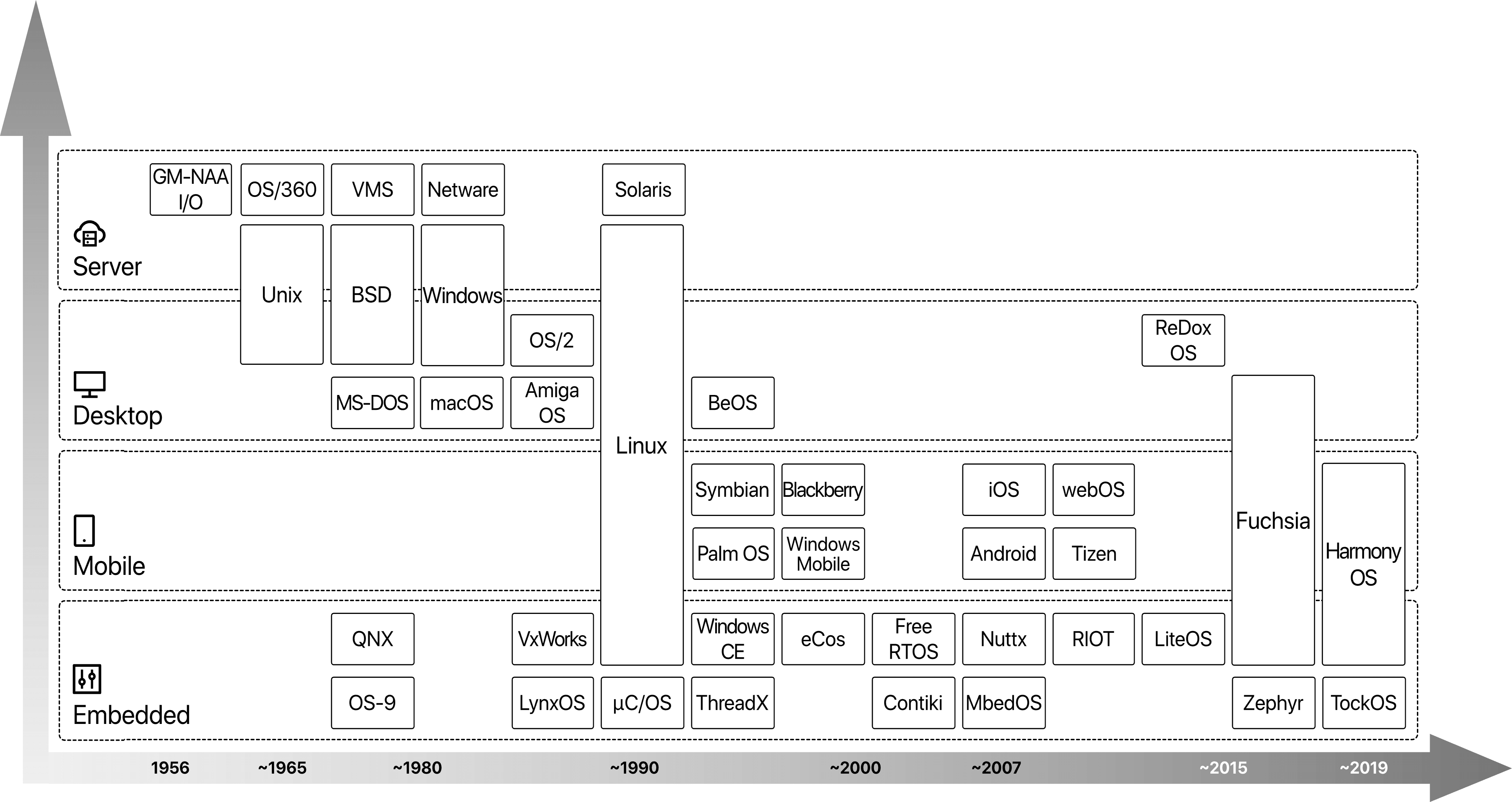}
\caption{\textbf{Historical Timeline of Operating Systems Across Different Device Categories.} This figure provides a comprehensive overview of operating systems across server, desktop, mobile, and embedded categories over time, spanning from the early days of computing in the mid-20th century up until recent years. }
\label{Existing Released Operating Systems}
\end{figure*}

\section{Existing Released Operating Systems}
\label{sec:Existing OS}

\subsection{Embedded Operating Systems}

\subsubsection{Introduction}
Embedded operating systems are specialized software designed to serve the unique needs of embedded systems, which are prevalent in various applications ranging from household appliances and industrial control devices to automotive electronics and aerospace instrumentation. These OSs are characterized by their real-time capabilities, enabling them to respond to external events within predetermined timeframes. They operate under constrained resource environments, necessitating efficient execution within limited memory and storage spaces. They are also highly customizable, allowing them to be tailored to fit specific hardware platforms and prioritize stability and reliability, given their operation in harsh and variable conditions. 

\subsubsection{Characteristics}

Embedded operating systems are distinguished by their stringent real-time constraints, mandating that they respond to stimuli within a precise temporal framework. This is particularly critical in safety-critical applications such as automotive systems, where latency can have dire consequences. As such, the design of embedded operating systems is optimized to ensure timely task processing unaffected by the overhead of concurrent operations.

In addition to real-time performance, these operating systems face challenges due to the limitations of resources. Many embedded devices, constrained by cost, size, or power consumption, are equipped with low-performance processors and have restricted memory and storage capacities. This necessitates that both the operating system and the applications running atop it be lean and efficient, capable of functioning smoothly on such modest hardware. Moreover, the hardware configuration for a single embedded device is often unique, demanding that the operating system be tailored to support only a specific set of hardware, thus reducing the demand on system resources.

Furthermore, embedded operating systems emphasize a high degree of customizability. This is primarily due to the significant variance in hardware architecture and interface requirements among embedded devices. Embedded operating systems offer a flexible, modular structure to accommodate these differences, enabling developers to select and configure the necessary functional modules based on the specific application scenario. This flexibility facilitates adaptation to diverse hardware products' needs, minimizes code volume, and enhances overall system performance.

Lastly, stability and reliability are critical characteristics of embedded operating systems that cannot be overlooked. As many embedded systems operate unattended and often under extreme conditions such as significant temperature fluctuations, strong electromagnetic interference in industrial settings, or remote areas that are difficult to maintain, the operating system must be capable of long-term fault-free operation and possess a certain degree of fault tolerance. Consequently, the design of embedded operating systems typically incorporates a suite of mechanisms to ensure reliable system operation, including error detection and correction, fault isolation, and other techniques. These features collectively ensure that embedded operating systems are up to the complex tasks of their respective domains.

\subsubsection{Development History}

\begin{figure*}[!t]
\centering
\includegraphics[draft=false, width=\textwidth]{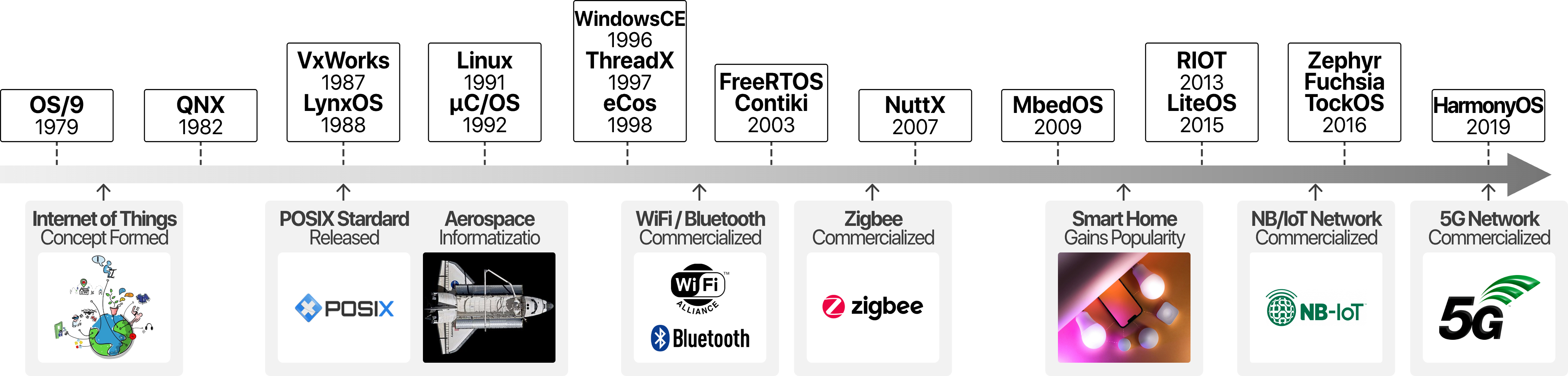}
\caption{\textbf{Chronological Development of Embedded Operating Systems and Associated Technological Milestones.} This figure chronicles the evolution of embedded operating systems from their origins in the late 20th century to contemporary platforms, alongside pivotal technological advancements that have driven innovation in this domain. }
\label{Development of Embedded Operating Systems}
\end{figure*}

\begin{table*}[!htpb]
\centering
\caption{Embedded OS Characteristic}
\renewcommand{\arraystretch}{1.8} 
\begin{tabular}{|c|c|c|c|}  
\hline
\textbf{OS} & \textbf{ISA Supported} & \textbf{Kernel Architecture} & \textbf{Open Sourced}  \\ \hline

\rule{0pt}{16pt} \makecell{\textbf{OS-9} \vspace{1.5pt}}& \makecell{X86, ARM, MIPS, PowerPC, Motorola 6809, \\Motorola 680x0, ColdFire, SuperH \vspace{1.5pt}}  & \makecell{Microkernel \vspace{1.5pt}} & \makecell{No \vspace{1.5pt}} \\ \hline

\textbf{QNX}  & \makecell{X86, ARM}  & \makecell{Microkernel} & \makecell{No} \\ \hline

\textbf{VxWorks} & \makecell {X86, ARM, RISC-V, MIPS, PowerPC, SuperH} & \makecell{Monolithic}  & \makecell{No}  \\ \hline

\textbf{LynxOS} & \makecell{Motorola 68010, X86, ARM, PowerPC}  & \makecell{Monolithic} & \makecell{No} \\ \hline

\rule{0pt}{21pt} \makecell{\textbf{Linux} \vspace{2.5pt}} & \makecell{Alpha, ARC, ARM, C-SKY, Hexagon,  LoongArch, MIPS,  \\Motorola 68000, MicroBlaze, Nios II,  OpenRISC, PA-RISC, \\ PowerPC,  RISC-V, S/390, SuperH, SPARC, x86, Xtensa \vspace{2.5pt}} & \makecell{Monolithic \vspace{2.5pt}} & \makecell{Yes \vspace{2.5pt}} \\ \hline

\textbf{µC/OS} & \makecell{PIC, AVR, ARM, MIPS, ColdFire, X86, RISC-V, MSP430} & \makecell{Microkernel} & \makecell{Partial} \\ \hline

\textbf{Windows CE} & \makecell{X86, ARM}  & \makecell{Hybrid} & \makecell{No} \\ \hline

\rule{0pt}{16pt} \makecell{\textbf{ThreadX} \vspace{1.5pt}} & \makecell{ARC, ARM, Blackfin, CEVA, C6x, MIPS, NXP, PIC, PowerPC, \\ RISC-V, RX, SuperH, SHARC, TI, V850, Xtensa, X86, Coldfire \vspace{1.5pt}}  & \makecell{Microkernel \vspace{1.5pt}} & \makecell{No \vspace{1.5pt}} \\ \hline

\rule{0pt}{16pt} \makecell{\textbf{eCos} \vspace{1.5pt}} & \makecell{ARM, CalmRISC, FR-V, X86, Matsushita AM3x,  Motorola 68000,\\ NEC V850, PowerPC,  SuperH, Hitachi H8,MIPS,  Nios II, SPARC \vspace{1.5pt}}  & \makecell{Hybrid \vspace{1.5pt}} & \makecell{Yes \vspace{1.5pt}} \\ \hline

\rule{0pt}{21pt} \makecell{\textbf{FreeRTOS} \vspace{2.5pt}} & \makecell{Robust Support including: ARM, MicroBlaze, RISC-V, PIC, \\ H8/S, SuperH, RX, 8052, ColdFire, V850, 78K0R, \\ Xtensa, Nios II,  FR,  RISC-V, X86, AVR,  MSP430 \vspace{2.5pt}}  & \makecell{Microkernel \vspace{2.5pt}} & \makecell{Yes \vspace{2.5pt}} \\ \hline

\textbf{Contiki} & \makecell{ARM, 8051, AVR, MSP430, NXP LPC, STM32, CC2538,CC26xx} & \makecell{Microkernel} & \makecell{Yes} \\ \hline

\textbf{NuttX} & \makecell{	ARM, AVR, HCS12, LM32, MIPS, RISC-V, SuperH, Xtensa, Z80} & \makecell{Monolithic} & \makecell{Yes} \\ \hline

\textbf{Mbed OS} & \makecell{ARM}  & \makecell{Microkernel} & \makecell{Yes} \\ \hline

\textbf{Zephyr} & \makecell{ARM, ARC, MIPS, NiosII, RISC-V, Xtensa, SPARC, X86}  & \makecell{Monolithic} & \makecell{No} \\ \hline

\textbf{Fuchsia} & \makecell{X86, ARM} & \makecell{Microkernel} & \makecell{Yes} \\ \hline

\textbf{TockOS} & \makecell{ARM, RISC-V}  & \makecell{Monolithic} & \makecell{Yes} \\ \hline

\textbf{HarmonyOS} & \makecell{ARM, X86, RISC-V} & \makecell{Hybrid} & \makecell{Partial} \\ \hline

\end{tabular}
\label{tab:embedded_os_metrics}
\end{table*}

With the rapid advancement of microprocessor technology, embedded operating systems have gradually become a critical component in meeting the needs of small computing devices and embedded systems. In 1979, Motorola introduced \href{https://en.wikipedia.org/wiki/OS-9}{OS-9}, a multitasking operating system designed for emerging small computers and embedded systems. OS-9, with its exceptional real-time performance and support for various hardware platforms, became an essential choice in early embedded applications. The birth of this system reflected the market's strong demand for efficient management and scheduling of limited computing resources at the time.

Entering the 1980s, the demand for system reliability and security significantly increased as the application scenarios of embedded systems became increasingly complex. In 1982, QNX\cite{hildebrand1992architectural} emerged as a real-time operating system with a microkernel architecture. Its kernel retained only the core functions required for operation, with most system components running in user space and inter-process communication achieved through a messaging mechanism. This design greatly enhanced system stability and security, marking a significant step in modular design and scalability of embedded operating systems to address more complex application scenarios.

As the informatization process in the aerospace and defense sectors accelerated, embedded operating systems' real-time processing capabilities and reliability attracted increasing attention. In 1987, Wind River Systems released VxWorks \cite{VxWorks}, an operating system specifically designed to meet real-time and high-reliability requirements. VxWorks supports multiple processor architectures and is highly customizable, adapting to specific hardware needs and thus gaining widespread application in the aviation, aerospace, and defense sectors.

However, early embedded systems faced significant challenges in cross-platform application compatibility due to the lack of unified standards. To address this issue, in 1988, Lynx Software Technologies introduced LynxOS\cite{singh1990lynxos}, a real-time operating system based on the POSIX standard. The system supports hard real-time task scheduling, making it particularly suitable for critical task systems with high reliability and security requirements. The introduction of LynxOS reflected the trend of embedded operating systems moving towards standardization and the industry's increased expectations for system security and reliability.

Entering the 1990s, the rise of the Internet and the proliferation of personal computers(PCs) propelled the development of the open-source software movement. In 1991, Linus Torvalds released the first version of Linux\cite{Linux}, which, although initially not designed for embedded systems, quickly became one of the important operating systems in the embedded field due to its openness, portability, and extensive community support. Concurrently, Jean J. Labrosse developed µC/OS, a real-time operating system specifically designed for embedded applications, which gained popularity for its lightweight nature, ease of portability, and excellent interrupt handling capabilities.

Entering the 21st century, the diversification trend of embedded operating systems became increasingly evident. In 1996, Microsoft released Windows CE\cite{levy1997windows}, an operating system designed for embedded devices, aiming to provide a user experience similar to desktop Windows. Due to its development environment being compatible with desktop Windows, Windows CE has found widespread application in industrial control devices and mobile devices.

With the rise of mobile Internet and the IoT, low power consumption and fast startup have become new requirements for embedded systems. To meet these needs, operating systems such as ThreadX\cite{ThreadX} (1997) and eCos\cite{domahidi2013ecos} (1998) were introduced. During this period, embedded operating systems had to satisfy real-time requirements while adapting to the resource-constrained characteristics of IoT devices.

The rapid development of IoT and edge computing has driven embedded operating systems further to penetrate fields such as sensor networks and smart homes. In the early 2000s, operating systems such as FreeRTOS\cite{barry2008freertos} (2003), Contiki\cite{dunkels2004contiki} (2003), NuttX\cite{Nuttx} (2007), and Mbed OS\cite{MbedOS} (2009) emerged. Most of these systems are designed for low-power microcontrollers and are specifically optimized for sensor networks and smart home applications.

Furthermore, with the increasing demand for device interconnectivity, some embedded operating systems have placed greater emphasis on cross-platform compatibility and data security. Major technology companies and open-source communities have introduced embedded operating systems, such as Google's Fuchsia\cite{Fuchsia} (2016), Huawei's HarmonyOS\cite{HarmonyOS} (2019), and TockOS\cite{levy2017tock} (2016), which specifically addresses security and memory safety in resource-constrained environments. These efforts aim to secure a significant position in the IoT and smart device market.

Overall, the development history of embedded operating systems demonstrates how technological advancements continuously adapt to and drive changes in market demands. From early simple task scheduling to today's complex distributed systems, each stage of development of embedded operating systems has solved the technical challenges of their time within a specific historical context, laying a solid foundation for future progress.

\subsection{Mobile Operating Systems}

\subsubsection{Introduction}
Mobile operating systems are designed for smartphones, tablet computers, and other portable devices. They provide users with a user-friendly interface and support a variety of basic and advanced functionalities. With the proliferation of mobile devices worldwide, mobile operating systems have become integral to modern digital life. They enable people to access the Internet anytime and anywhere and offer rich tools for entertainment, social interaction, work, and learning.

\begin{figure*}[!t]
\centering
\includegraphics[width=\textwidth, keepaspectratio, draft=false]{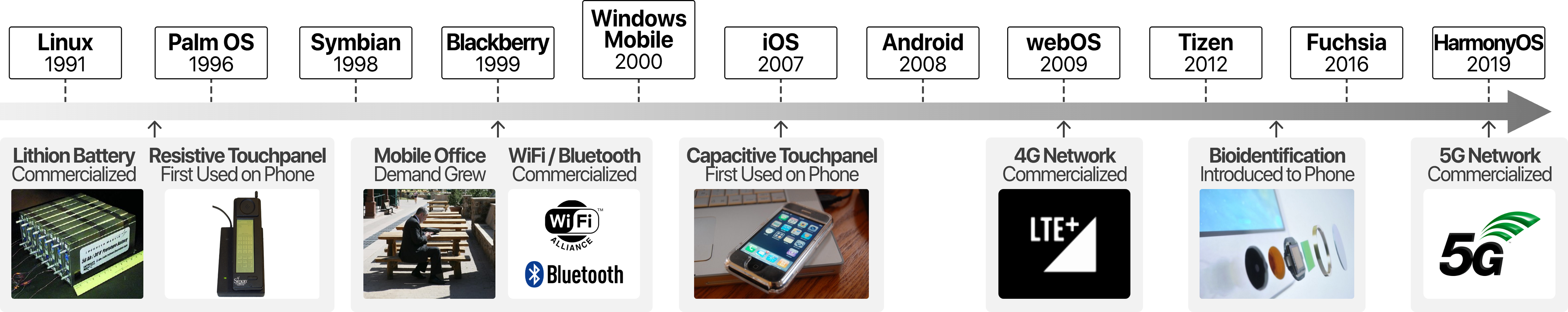}
\caption{\textbf{Historical Timeline of Mobile Operating System Evolution and Key Technological Advancements.} This figure presents a chronological timeline of the development of mobile operating systems from 1991 to 2019, along with significant technological milestones that contributed to their advancement. The timeline highlights the introduction of various mobile OSes. Additionally, it underscores critical innovations.}
\label{Timeline of Mobile Operating System Evolution}
\end{figure*}

\subsubsection{Characteristics}

Mobile operating systems are typically configured with a focus on high performance and low power consumption to meet portability demands and extended use. In terms of processing capabilities, modern smartphones and tablet computers predominantly utilize multi-core processors based on the ARM architecture, with clock speeds exceeding 1 GHz. Common designs include quad-core, octa-core, and even more cores, providing ample computational power for multitasking and high-performance applications. Regarding memory capacity, current mobile devices typically have at least 2GB to 16GB or more RAM to ensure a smooth operational experience and rapid application switching. Storage-wise, built-in storage capacities start at 32GB, with high-end devices offering up to 512GB or even 1TB of storage space, and some devices support storage expansion via microSD cards. In terms of peripheral chip functionality, mobile devices integrate a wealth of connectivity and sensor technologies, including Wi-Fi, Bluetooth, NFC, GPS, cellular network modules, as well as accelerometers, gyroscopes, proximity sensors, ambient light sensors, etc. These components collectively provide users with comprehensive communication capabilities and intelligent sensing functions. Modern mobile devices commonly support high-definition displays, high-quality cameras, and biometric technologies such as fingerprint or facial recognition to enhance user experience and security.

One of the most notable features of mobile operating systems is their support for touchscreens. Compared to traditional keyboard and mouse input methods, touchscreens offer a more intuitive and direct mode of interaction, allowing users to complete complex tasks with simple gesture operations. Multi-touch technology further enhances this experience by allowing multiple fingers to operate on the screen simultaneously, enabling functions such as zooming and rotating, greatly enriching the user's interactive experience.

To fully leverage mobile devices' characteristics, mobile operating systems have built-in support for various hardware functionalities, including wireless network connections (Wi-Fi, Bluetooth, cellular networks), Global Positioning System (GPS), cameras, accelerometers, gyroscopes, etc. This integration transforms mobile devices from simple communication tools into versatile platforms that offer a wide range of functions, including communication, navigation, photography, and fitness tracking. With these hardware capabilities, users can easily access location information, take photos, and engage in video calls, among other activities.

Mobile operating systems typically provide an official app store through which users can download and install various applications. These applications span multiple domains, such as gaming, social networking, office productivity, and health, significantly enriching the functionality of mobile devices and offering users the possibility of personalized customization. The presence of app stores also fosters the development of the developer economy, encouraging more individuals to participate in the development of mobile applications.

Mobile operating systems emphasize battery management and energy-saving features, considering mobile devices' portability. By effectively managing system resources, such as Central Processing Unit (CPU) frequency modulation and background app restrictions, mobile operating systems can extend device usage time and reduce the need for users to recharge frequently. Additionally, these systems offer a range of power management tools to help users monitor and optimize their device's power consumption.

\subsubsection{Development History}

\begin{table*}[!htpb]
\centering
\caption{Mobile OS Characteristic}
\renewcommand{\arraystretch}{1.8} 
\begin{tabular}{|c|c|c|c|}  
\hline
\textbf{OS} & \textbf{ISA Supported} & \textbf{Kernel Architecture} & \textbf{Open Sourced}  \\ \hline

\rule{0pt}{21pt} \makecell{\textbf{Linux} \vspace{2.5pt}} & \makecell{Alpha, ARC, ARM, C-SKY, Hexagon,  LoongArch, MIPS,  \\Motorola 68000, MicroBlaze, Nios II,  OpenRISC, PA-RISC, \\ PowerPC,  RISC-V, S/390, SuperH, SPARC, x86, Xtensa \vspace{2.5pt}} & \makecell{Monolithic \vspace{2.5pt}} & \makecell{Yes \vspace{2.5pt}} \\ \hline

\makecell{\textbf{PalmOS} \\} & \makecell{ARM, MIPS} & \makecell{Microkernel} & \makecell{No} \\ \hline

\makecell{\textbf{Symbian} \\} & \makecell{ARM, X86} & \makecell{Microkernel} & \makecell{No} \\ \hline

\makecell{\textbf{Windows Mobile} \\} & \makecell{ARM} & \makecell{Hybrid} & \makecell{No} \\ \hline

\makecell{\textbf{iOS} \\} & \makecell{ARM} & \makecell{Hybrid} & \makecell{No} \\ \hline

\makecell{\textbf{Android} \\} & \makecell{ARM, X86, RISC-V} & \makecell{Monolithic} & \makecell{CFS} \\ \hline

\makecell{\textbf{webOS} \\} & \makecell{ARM, X86} & \makecell{Monolithic} & \makecell{Partial} \\ \hline

\makecell{\textbf{Tizen} \\} & \makecell{ARM, X86, MIPS} & \makecell{Hybrid} & \makecell{Yes} \\ \hline

\makecell{\textbf{Fuchsia} \\} & \makecell{ARM, X86} & \makecell{Microkernel} & \makecell{Yes} \\ \hline

\makecell{\textbf{HarmonyOS} \\} & \makecell{ARM, RISC-V, X86, Xtensa, C-SKY} & \makecell{Hybrid} & \makecell{Partial} \\ \hline

\end{tabular}
\label{tab:mobile_os_metrics}
\end{table*}

By the late 1990s, the hardware capabilities of mobile phones were continuously enhanced, particularly with improvements in processor performance, storage capacity, and battery technology, enabling mobile devices to run more complex operating systems. Concurrently, as user demands for mobile communication expanded from simple call functions to text messaging, email, and basic internet access, mobile phone manufacturers began to explore platforms suitable for multitasking and application development. In 1998, the Symbian\cite{Symbian} operating system emerged, a collaborative development by several mobile phone manufacturers aimed at providing unified support for high-end mobile phones. Symbian supported multiple hardware architectures and offered a rich set of Application Programming Interfaces (APIs), providing developers with a robust environment for application development. However, Symbian's relatively closed architecture struggled to adapt flexibly to the rapidly evolving touchscreen technology especially in applying multi-touch capabilities, and eventually got replaced by emerging smartphone operating systems.

With the increasing demand for mobile office solutions, Palm, Inc. introduced the Palm OS\cite{jones1995palm} in 1996, an operating system specifically designed for mobile devices, focusing on personal information management (PIM) and productivity. The Palm OS featured a user-friendly interface and efficient data synchronization capabilities, making it ideal for managing contacts, calendars, and to-do lists. Its lightweight design and long battery life were particularly appealing to business professionals who needed reliable and portable devices for on-the-go use. The inclusion of a stylus and a Graffiti handwriting recognition system allowed users to input data quickly and accurately, enhancing the overall user experience. The Palm OS achieved significant success in the early mobile market, establishing itself as a leading platform for personal digital assistants (PDAs) and laying the groundwork for future mobile operating systems.

With the rise of mobile office needs for business, RIM (Research In Motion) launched the BlackBerry OS\cite{BlackBerry_OS} in 1999, an operating system specifically designed for enterprise users, focusing on meeting email and instant messaging needs. The BlackBerry OS combined physical keyboards with the BlackBerry Messenger service, achieving significant success in the enterprise market. Its hardware-centric design allowed business users to handle information quickly and securely.

Entering the early 2000s, with the proliferation of the internet, mobile devices were no longer just communication tools but had become multifunctional information terminals, with users expecting to perform more complex computing tasks through their mobile phones. In 2000, Microsoft introduced Windows Mobile, an operating system based on the Windows CE core, attempting to transplant the desktop PC experience to mobile phones. Windows Mobile\cite{Windows_Mobile} supported touchscreen operations and came with built-in productivity tools like Office suites, meeting the mobile office needs of some users. However, Windows Mobile did not achieve widespread success due to its complex interface design and high hardware requirements, which resulted in poor performance on resource-limited mobile devices.

With the maturation of touchscreen technology and significant improvements in processor performance, mobile device interaction methods and functionalities underwent revolutionary changes. Apple's iOS\cite{ios} operating system, released in 2007, completely transformed how mobile devices were used. iOS replaced the physical keyboard with a full touchscreen interface, allowing users to perform intuitive operations through multi-touch, greatly enhancing the user experience. Additionally, iOS introduced the App Store as an application distribution platform, allowing third-party developers to create applications for this platform, establishing an ecosystem that significantly promoted the prosperity of mobile applications.

Following Apple's lead in the market, mobile device manufacturers gradually recognized the necessity of software openness and flexibility. In 2008, Google launched the Android\cite{gilski2015android} operating system based on the Linux kernel. Android's open architecture allowed device manufacturers to deeply customize the system and supported a variety of hardware configurations, from high-end smartphones to entry-level devices, enabling Android to dominate the market quickly. Android not only attracted a large number of developers through its open ecosystem, but Google Play Store also provided users with a vast selection of applications, meeting diverse user needs.

In 2012, the Tizen Association, led by Samsung and Intel, introduced Tizen\cite{vashisht2014study}, another Linux-based operating system aimed at a wide range of devices, including smartphones, tablets, and smart TVs. Tizen was designed to offer a highly customizable and flexible platform, supporting both open-source and proprietary applications. It featured a robust multimedia framework and strong integration with web technologies, aligning with the growing trend of web-centric computing. Tizen aimed to provide a seamless user experience across multiple devices, emphasizing connectivity and interoperability. However, despite its technical merits, Tizen faced challenges in building a substantial app ecosystem and gaining widespread adoption, primarily due to the dominance of Android and iOS in the mobile market. Nonetheless, Tizen continues to be developed and is used in certain niche markets, such as wearables and smart home devices.

With the rapid development of IoT and AI technologies, user demands for mobile operating systems gradually expanded to include cross-device connectivity and intelligent interaction. In 2016, Google first revealed the Fuchsia\cite{Fuchsia} operating system, a next-generation operating system based on a microkernel aimed at providing a unified cross-platform solution that can run on everything from mobile phones to smart home devices. Fuchsia's design is well-prepared for future AI and natural user interfaces, such as voice and gesture interactions.

In response to the need for IoT and multi-device collaboration, Huawei launched HarmonyOS\cite{HarmonyOS} in 2019. This distributed operating system aims to break through the limitations of single devices to achieve seamless collaboration between diverse devices such as smartphones, tablets, smartwatches, and smart home devices. HarmonyOS aims to provide users with a unified service experience, allowing them to switch and interact freely within the same ecosystem, regardless of device performance differences. The launch of this system signifies that mobile operating systems are gradually moving towards the development of a full-scenario smart living direction.

\subsection{Desktop Operating Systems}

\begin{figure*}[!t]
\centering
\includegraphics[draft=false, width=\textwidth]{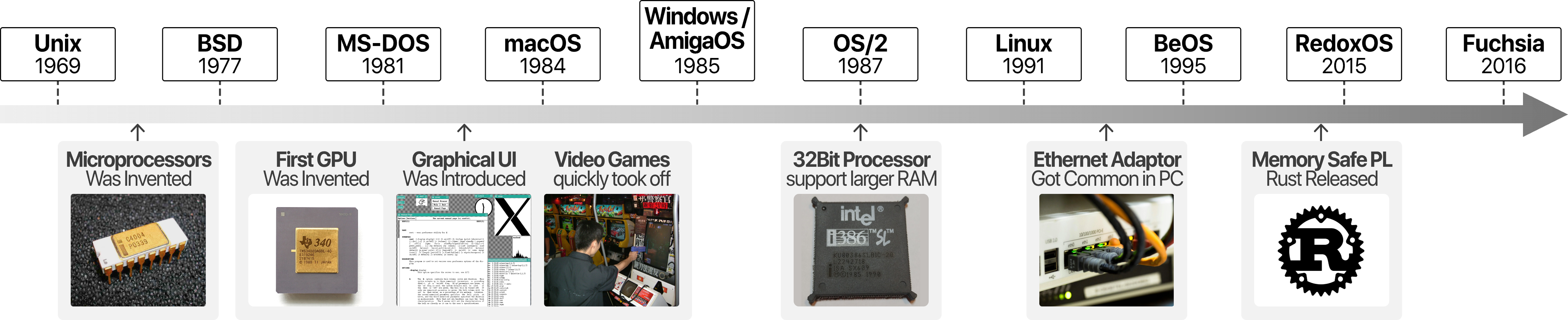}
\caption{\textbf{Chronological Development of Desktop Operating Systems and Associated Technological Milestones.} This timeline traces the historical progression of desktop operating systems from their inception to modern times, alongside significant technological advancements that influenced their evolution.}
\label{Development of Desktop Operating Systems}
\end{figure*}

\subsubsection{Introduction}
A desktop operating system refers to a category of software designed to provide management and control functionalities for PCs, laptops, and workstations. It offers users a Graphical User Interface (GUI) that facilitates interaction with computer hardware through mouse clicks and keyboard inputs, enabling the execution of various operations. Since the 1980s, desktop operating systems have become the core of personal computing, profoundly altering people's working methods and lifestyles.

\subsubsection{Characteristics}
The configuration of devices running desktop operating systems typically aims at providing robust computational capabilities and rich multimedia experiences. In terms of processors, modern desktops and laptops predominantly utilize multi-core processors based on x86 or ARM architectures, with clock speeds ranging from 2GHz to over 5GHz and core counts varying from dual-core to sixteen cores or more to support complex applications and multitasking. Regarding memory capacity, contemporary desktop devices usually have at least 8GB to 64GB or more of RAM to ensure efficient data processing and a smooth user experience. For storage, devices commonly adopt Solid State Drives (SSDs) as primary storage media, with capacities ranging from 128GB to several terabytes; some devices also incorporate Hard Disk Drives (HDDs) to offer additional storage space. In peripheral functions, desktop devices integrate high-performance Graphics Processing Units (GPUs) to support graphics-intensive applications such as gaming and professional design software. They also feature high-speed network interfaces (such as Gigabit Ethernet or Wi-Fi 6), USB 3.0 or higher ports, Thunderbolt, HDMI, and audio interfaces to enable rapid data transfer and multimedia connectivity. Moreover, modern desktop devices support multi-monitor outputs, high-quality audio systems, and advanced cooling designs to ensure stable operation and an excellent user experience.

The primary purpose of desktop operating systems is to provide users with a friendly interface to execute a variety of complex tasks, ranging from daily home entertainment to professional office applications. By offering a unified interface standard, desktop operating systems simplify the learning curve, allowing even novice computer users to get started quickly. Additionally, desktop operating systems support multitasking, enabling users to run multiple applications simultaneously without worrying about system crashes or performance degradation. This capability is particularly crucial for users who need to handle multiple tasks concurrently.

Support for a multi-user environment is an important feature of desktop operating systems. In certain scenarios, especially in corporate or educational settings, a single computer may be shared by multiple users. To protect the security and privacy of personal data, desktop operating systems provide account management systems that allow each user to have their own login credentials and set different permission levels. This way, each user can work freely within their own environment without affecting the data of other users.

In addition to basic functionalities, most desktop operating systems pre-install common applications such as office suites, web browsers, and media players. These applications are intended to assist users in completing routine tasks more efficiently, such as editing documents, browsing the web, and playing music or videos. Furthermore, desktop operating systems support various methods for installing and managing applications, including application stores, software package management systems, and sideloading, enabling users to extend their computers' functionality easily.

User experience and ease of use are critical metrics for evaluating the quality of a desktop operating system. Over time, the design of operating systems has increasingly emphasized humanization, striving to reduce the learning cost for users and enhance efficiency. Modern desktop operating systems typically feature intuitive user interfaces, clear icons and menu layouts, and powerful search functions. Additionally, to cater to the preferences of different users, operating systems permit customization of visual elements such as desktop backgrounds, window colors, and font sizes.

Over the past few decades, desktop operating systems have undergone tremendous transformations and developments. From early command-line interfaces (CLI) to today's GUIs, from single-user systems to multi-user systems, and from supporting only a few languages to accommodating multiple languages globally, every change has significantly enhanced the user experience. Although the importance of desktop operating systems seems to have diminished with the proliferation of mobile devices, they remain indispensable for users requiring high-performance computing and professional applications.

\subsubsection{Development History}

\begin{table*}
\centering
\caption{Desktop OS Characteristic}
\renewcommand{\arraystretch}{1.8} 
\begin{tabular}{|c|c|c|c|} 
\hline
\textbf{OS} & \textbf{ISA Supported} & \textbf{Kernel Architecture} & \textbf{Open Sourced}  \\ \hline

\rule{0pt}{16pt} \makecell{\textbf{Unix} \vspace{2pt}} & \makecell{Varies, including: X86, ARM, MIPS, PowerPC, SPARC \vspace{2pt}} & \makecell{Varies: Monolithic,\\Microkernel, Hybrid \vspace{2pt}} & \makecell{No \vspace{2pt}} \\ \hline

\textbf{MS-DOS} & X86 & Monolithic & No \\ \hline

\textbf{macOS} & X86, ARM & Hybrid & No \\ \hline

\textbf{Windows} & X86, ARM & Hybrid & No \\ \hline

\textbf{AmigaOS} & Motorola 68000, PowerPC & Microkernel & No \\ \hline

\textbf{OS/2} & X86, PowerPC & Hybrid & No \\ \hline

\rule{0pt}{21pt} \makecell{\textbf{Linux} \vspace{2.5pt}} & \makecell{Alpha, ARC, ARM, C-SKY, Hexagon,  LoongArch, MIPS,  \\Motorola 68000, MicroBlaze, Nios II,  OpenRISC, PA-RISC, \\ PowerPC,  RISC-V, S/390, SuperH, SPARC, x86, Xtensa \vspace{2.5pt}} & \makecell{Monolithic \vspace{2.5pt}} & \makecell{Yes \vspace{2.5pt}} \\ \hline

\textbf{BSD} & \makecell{Varies, including: X86, ARM, MIPS, PowerPC, RISC-V, SPARC} & Monolithic & Yes \\ \hline

\textbf{BeOS} & PowerPC, X86 & Hybrid & No \\ \hline

\textbf{RedoxOS} & X86 & Microkernel & Yes \\ \hline

\textbf{Fuchsia} & ARM, X86 & Microkernel & Yes \\ \hline

\end{tabular}
\label{tab:desktop_os_metrics}
\end{table*}

The evolution of desktop operating systems reflects continuous advancements in hardware technology and changes in user demands. From early text-based processing to later GUIs to the widespread application of multimedia and the Internet in the 21st century, desktop operating systems have evolved continuously to adapt to changing computing environments.

By the end of 1960s, with improvements in computer hardware performance and the demand for multi-user systems, operating system design began to mature. Unix\cite{Unix}, born in 1969 at Bell Labs, established foundational principles through its hierarchical file system and multitasking capabilities. Unix's modularity, simplicity, and portability philosophy allowed it to run across different hardware platforms. By the 1980s, Unix's influence had expanded, and many subsequent operating systems directly or indirectly adopted Unix's design concepts.

Building on the success of Unix, the Berkeley Software Distribution (BSD) \cite{quarterman19854} emerged in 1977 at the University of California, Berkeley. BSD extended Unix by adding features such as the virtual memory system, TCP/IP networking support, and the Berkeley sockets API, which became essential for internet communication. These enhancements significantly improved the performance and functionality of Unix-based systems, making them more suitable for academic and research environments. BSD's emphasis on open-source software and collaborative development fostered a vibrant community of developers and researchers, contributing to numerous innovations and improvements. By the 1980s, BSD had become a cornerstone in the development of networked computing, influencing the creation of several derivatives and contributing to the widespread adoption of Unix-like systems in both academia and industry.

In the 1980s, breakthroughs in microprocessor technology enabled PCs to reach a broader market. IBM launched its first PC in 1981, powered by Intel’s 8088 processor, marking the beginning of the personal computing era. Accompanying this hardware was Microsoft’s DOS\cite{paterson1983inside}, a single-user, single-task command-line operating system that allowed direct interaction with hardware via a command prompt. Despite its limited functionality, MS-DOS laid the groundwork for more sophisticated graphical operating systems.

By the mid-1980s, processor speed and memory capacity advancements improved computer graphics processing abilities. The demand for more intuitive user interactions drove the development of GUIs. In 1984, Apple introduced the Macintosh computer, equipped with System Software (later renamed macOS\cite{macOS}), the first widely used commercial operating system featuring a GUI. macOS utilized desktop metaphors, window management, menus, and a mouse to facilitate more intuitive and convenient interaction between users and computers. The introduction of GUI significantly lowered the barrier to computer use and transformed computers into tools for creativity, design, and multimedia work.

Around the same time, Microsoft released Windows\cite{Windows} in 1985 as an extension to DOS. The initial versions of Windows were graphical environments built on top of DOS. While still technically dependent on the command-line operating system, these versions improved user experience through simple graphical interfaces. As computing hardware evolved, particularly with the advent of GPU, graphical interfaces became mainstream in desktop operating systems.

In the 1990s, significant increases in processor performance made multitasking and multimedia computing core requirements of desktop computing. Microsoft’s Windows NT\cite{WindowsNT}, released in 1993, featured a more stable and modular kernel design to support multiprocessor systems and advanced memory management features. Windows NT introduced advanced networking support (e.g., TCP/IP protocols), aligning with the growing trend of internet usage and becoming an ideal choice for enterprise servers and workstations. Simultaneously, Microsoft continued enhancing user interfaces and multimedia support through Windows 95 and Windows 98 updates.

During this period, operating systems like AmigaOS\cite{iamigaos} (1985) promoted the development of multimedia creation and entertainment applications. AmigaOS’s multitasking and advanced graphic and sound processing capabilities distinguished it in game development and multimedia creation, reflecting users’ expectations for entertainment features in computing devices.

With the rise of the Internet, open-source software gained widespread support. In 1991, Linus Torvalds released the Linux\cite{Linux} kernel, based on Unix's design principles, offering personal users a free, open-source, and highly customizable operating system. Linux rapidly succeeded in server environments and gained a foothold in desktop computing through various distributions such as Debian and Red Hat. Its success was mainly due to its openness and community support, driving innovation in software development during the Internet age.

Similarly, Unix-based open-source operating systems like FreeBSD\cite{quarterman19854} (1993) emerged, inheriting Unix’s stability and security while demonstrating superior performance in server and embedded systems.

Entering the 21st century, with the popularization of multi-core processors, SSD storage devices, and high-speed networking technologies, users placed higher demands on the performance and security of desktop operating systems. The emergence of operating systems like BeOS\cite{BeOS} in the early 2000s reflected the market’s need for multimedia processing capabilities. Although BeOS failed to gain widespread adoption due to poor marketing, its optimization for multiprocessor support and real-time processing represented innovative directions in multimedia computing.

In recent years, experimental operating systems like Redox OS\cite{ReDox} (2015), written entirely in Rust, have explored more modern and secure operating system architectures. Redox’s appearance reflects ongoing exploration in the field of operating systems towards higher security and performance.

With the rapid development of cloud computing and the IoT, the role of operating systems has changed. In 2016, Google introduced the Fuchsia\cite{Fuchsia} project, aiming to build a cross-platform operating system to support wide-ranging applications from smartphones to IoT devices. Fuchsia’s microkernel design reflects Google’s vision for future distributed computing environments.

\subsection{Server Operating Systems}

\begin{figure*}[!t]
\centering
\includegraphics[draft=false, width=\textwidth]{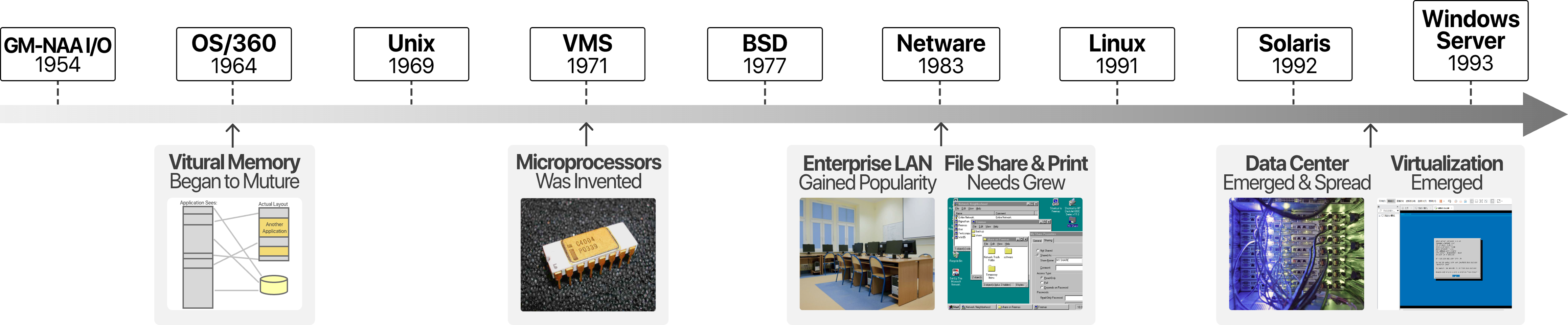}
\caption{\textbf{Timeline of Key Developments in Server Operating Systems and Related Technologies.} This timeline illustrates the chronological sequence of pivotal events and innovations in the realm of server operating systems and associated technologies, spanning from the introduction of virtual memory in the late 1950s to the emergence of virtualization in the early 21st century. }
\label{Timeline of Key Developments in Server Operating Systems and Related Technologies}
\end{figure*}

\subsubsection{Introduction}
Server operating systems are specialized software designed to run on server hardware. They are primarily aimed at providing reliable operational environments for businesses and network services. These operating systems manage underlying hardware resources and optimize network communication and data processing capabilities to meet the concurrent requests of numerous users and applications. Typically deployed in data centers, cloud environments, and wherever high-performance computing and storage services are required, server operating systems are pivotal in supporting mission-critical applications and services.

\subsubsection{Characteristics}

The configuration of devices running server operating systems is centered around high availability, reliability, and robust computational capabilities. In terms of processors, servers predominantly employ high-performance multi-core processors based on X86 architecture, featuring dozens to hundreds of cores, with clock speeds typically ranging from 2GHz to 5GHz, and support for hyper-threading technology to enhance concurrent processing capabilities. In recent years, ARM architecture processors have also begun making inroads into the server space. They are characterized by low power consumption and high density and suitable for large-scale cloud service deployments. 

Regarding memory capacity, server equipment often comes equipped with tens of gigabytes to several terabytes of RAM, supporting extensive data caching and multitasking to ensure the efficient operation of applications. For storage, servers utilize high-speed SSDs alongside HDDs as primary storage media, employing RAID technology to provide storage tiering, data redundancy, and fault tolerance, with storage capacities ranging from hundreds of gigabytes to petabytes.

Modern servers frequently include accelerators like GPU, TPU, or Data Processing Unit (DPU), which provide additional computational power to meet specific computational needs, such as machine learning (ML), high-performance computing, and graphics rendering. These accelerators can significantly enhance the processing speed of specific tasks, particularly when handling large datasets or executing complex algorithms. 

Peripheral server chipsets include high-performance Network Interface Cards (NICs) supporting 10GbE and faster network connections, enabling rapid data transmission. Additionally, servers integrate dedicated management processors (like IPMI or DRAC) for remote monitoring and system state management and support hot-swappable hard drives and redundant power supplies to ensure high availability and stability. These features collectively form the backbone of modern servers' powerful computing and data processing capabilities, enabling them to support complex enterprise applications and services.

One notable characteristic of server operating systems is their network and server hardware performance optimization. To address the demands of large-scale data transmission and processing, these operating systems are typically equipped with efficient network stacks capable of handling high bandwidth and low latency requirements. Moreover, server operating systems are optimized for different hardware architectures to ensure optimal performance on multi-core processors, large memory volumes, and high-speed storage devices. Such optimizations ensure servers respond promptly to client requests and process data streams efficiently.

High concurrency handling and support for multitasking are among the core strengths of server operating systems. In typical server environments, thousands of users might access a single application or service simultaneously, necessitating that the operating system effectively allocates resources to ensure the timely processing of each request. Server operating systems generally adopt multi-threaded or multi-process models and advanced scheduling algorithms to support massive concurrent connections and tasks. This capability allows servers to handle diverse requests from various sources without compromising overall performance.

Security is an essential characteristic of server operating systems that cannot be overlooked. Since servers often carry sensitive information and critical business processes, they are prime cyberattack targets. To protect data and system integrity, server operating systems integrate multi-layer security measures, including firewalls, intrusion detection systems, data encryption techniques, and regular security auditing functions. Additionally, many server operating systems provide fine-grained access control mechanisms, allowing administrators to restrict access to specific resources based on user roles and permissions.

Data management functions are also vital components of server operating systems. Whether database services, file sharing, or email systems, a reliable mechanism for storing and backing up data is necessary. Server operating systems typically include advanced file systems supporting RAID configurations to enhance data redundancy and recovery capabilities. They also provide automated backup tools to help administrators regularly back up important data and restore it quickly when needed, minimizing the risk of data loss.

Finally, server operating systems are designed to support 24/7 operations. This means they must possess extremely high stability and reliability to ensure continuous service provision under any circumstances. To achieve this, server operating systems are typically equipped with automatic fault recovery features that detect and fix issues at both the hardware and software levels, supporting hot-swapping technology to replace faulty components without downtime. Furthermore, these operating systems offer detailed monitoring and logging functionalities, helping administrators identify potential problems promptly and take preventive measures.

\subsubsection{Development History}

\begin{table*}[!htpb]
\centering
\caption{Server OS Characteristic}
\renewcommand{\arraystretch}{1.8} 
\begin{tabular}{|c|c|c|c|}  
\hline
\textbf{OS} & \textbf{ISA Supported} & \textbf{Kernel Architecture} & \textbf{Open Sourced}  \\ \hline

\makecell{\textbf{GM-NAA I/O} \\} & \makecell{IBM 704} & \makecell{Monolithic} & \makecell{No} \\ \hline

\rule{0pt}{16pt} \makecell{\textbf{OS/360} \vspace{1.5pt}} & \makecell{S/360, S/370 \vspace{1.5pt}} & \makecell{Varies: Monolithic, \\ Microkernel, Hybrid \vspace{1.5pt}} & \makecell{No \vspace{1.5pt}} \\ \hline

\rule{0pt}{16pt} \makecell{\textbf{Unix} \vspace{1.5pt}} & \makecell{Varies, including: X86, ARM, MIPS, PowerPC, SPARC \vspace{1.5pt}} & \makecell{Varies: Monolithic,\\Microkernel, Hybrid \vspace{1.5pt}} & \makecell{No \vspace{1.5pt}} \\ \hline

\makecell{\textbf{VMS} \\} & \makecell{VAX, Alpha, Itanium, X86} & \makecell{Monolithic} & \makecell{Yes} \\ \hline

\makecell{\textbf{BSD} \\} & \makecell{Varies, including: X86, ARM, MIPS, PowerPC, RISC-V, SPARC} & \makecell{Monolithic} & \makecell{Yes} \\ \hline

\makecell{\textbf{NetWare} \\} & \makecell{X86, MIPS, PowerPC, SPARC, Alpha} & \makecell{Hybrid} & \makecell{No} \\ \hline

\rule{0pt}{21pt} \makecell{\textbf{Linux} \vspace{2.5pt}} & \makecell{Alpha, ARC, ARM, C-SKY, Hexagon,  LoongArch, MIPS,  \\Motorola 68000, MicroBlaze, Nios II,  OpenRISC, PA-RISC, \\ PowerPC,  RISC-V, S/390, SuperH, SPARC, x86, Xtensa \vspace{2.5pt}} & \makecell{Monolithic \vspace{2.5pt}} & \makecell{Yes \vspace{2.5pt}} \\ \hline

\makecell{\textbf{Solaris} \\} & \makecell{X86, SPARC} & \makecell{Monolithic} & \makecell{No} \\ \hline

\makecell{\textbf{Windows Server} \\} & \makecell{X86, ARM} & \makecell{Hybrid} & \makecell{No} \\ \hline

\end{tabular}
\label{tab:server_os_metrics}
\end{table*}

The evolution of server operating systems closely reflects advancements in hardware technology and the growing demands of enterprise computing. From the minicomputers of the 1960s to today's cloud computing and large-scale data centers, the development of server operating systems mirrors the rapid increase in computing power and supports the thriving growth of critical business and network services.

In the 1950s, one of the earliest computing systems to introduce input/output (I/O) operations was the GM-NAA I/O, developed by General Motors and North American Aviation for the IBM 701 computer. GM-NAA I/O, created in 1956, is considered one of the first operating systems. It was designed to streamline the execution of programs by automating the loading, executing, and outputting processes of the 701 computer, marking a significant step in automating early computer operations. GM-NAA I/O introduced key features such as a rudimentary form of scheduling, which allowed it to manage a sequence of jobs and handle basic I/O operations, which were crucial for the time. This pioneering system laid the groundwork for more complex and advanced operating systems, highlighting the importance of efficient I/O management and job scheduling in managing computational tasks. The foundational concepts of GM-NAA I/O, such as job scheduling and I/O handling, influenced the design of later operating systems and marked the beginning of operating system development in computing history.

In the 1960s, the development of server operating systems began to take shape with significant advancements in computing hardware and the increasing demand for multi-user, multitasking systems from enterprises and research institutions. One of the earliest and most influential systems was IBM's OS/360\cite{ii1972360}, introduced in 1964. OS/360 was designed to support a wide range of IBM mainframe computers, providing a unified and comprehensive operating environment. Its key features included robust job scheduling, advanced I/O management, and support for multiple programming languages. OS/360's ability to manage complex workloads and large datasets made it a cornerstone for enterprise computing, setting standards for reliability and efficiency.

In the 1970s, server operating systems continued to evolve. Unix\cite{Unix}, a pioneering operating system developed in 1971 by Ken Thompson and Dennis Ritchie at AT\&T Bell Labs, was written in C language, making it more portable across different hardware platforms. Unix's layered structure and modular design significantly increased the flexibility of operating systems and introduced features such as file systems, process management, and virtual memory. Most importantly, Unix pioneered multi-user and multitasking capabilities, meeting the urgent need of enterprises and academic institutions for shared computing resources. With the progress in hardware, Unix became the mainstream operating system for mainframes and minicomputers, laying a solid foundation for later server operating systems.

As we entered the late 1970s, virtual memory technology matured, enabling servers to manage memory resources more efficiently. DEC's VMS\cite{VMS} (Virtual Memory System), introduced in 1977, was a landmark achievement in virtual memory technology. Explicitly designed for DEC's VAX series of minicomputers, VMS provided a stable multi-user environment capable of managing a large number of programs and data efficiently. Beyond enhancing memory management, VMS supported network communication protocols and file systems, gradually promoting the transition of enterprises from single-system environments to distributed computing environments. VMS was widely applied in government, research institutions, and large enterprises, occupying a significant position in the server market then.

In the 1980s, Local Area Network (LAN) technology matured, prompting enterprises to build internal networks to meet the needs of file sharing and printing services. To support server operations within LANs, Novell released NetWare\cite{NetwareOverview} in 1983, marking the gradual integration of network services as a critical function of server operating systems. With its advanced file and print services, multi-user concurrent access, and robust network management capabilities, NetWare quickly became the preferred platform for building enterprise LANs. With the development of Ethernet and TCP/IP protocols, NetWare facilitated the popularization of LAN technology in enterprises and laid the technical foundation for future network operating systems.

With the leap in computer hardware performance in the 1990s, especially the emergence of multiprocessor systems, enterprises demanded higher performance computing and more complex network management. In 1993, Microsoft released Windows NT 3.1, the first operating system based on the NT kernel, explicitly designed for workstation and server environments. Windows NT supported multiprocessor architectures, enhanced memory management, and advanced network functionalities, meeting the requirements of enterprises for computing performance, system stability, and security. The modular design of the Windows NT kernel improved system scalability and allowed enterprises to deploy server roles according to their needs, such as file servers, database servers, and application servers.

During the same period, Sun Microsystems introduced Solaris\cite{Solaris}, a Unix-based operating system designed for high-performance servers. Supporting multiple hardware platforms, Solaris offered powerful network functionalities and security features. By introducing scalable thread and storage management systems, Solaris excelled in large-scale computing tasks and distributed environments, becoming the preferred choice in industries with high-performance computing needs, such as finance and telecommunications.

In the mid-to-late 1990s, with the rise of the open-source software movement, the development model of operating systems fundamentally changed. The release of the Linux\cite{Linux} kernel in 1991 marked the rise of open-source operating systems. Based on the Unix kernel, Linux's openness and modular design allowed it to rapidly develop within the global developer community, achieving significant success, particularly in server operating systems. Linux provided flexible multitasking, outstanding network performance, and multi-user support, excelling especially in Web servers, database servers, and supercomputing. Its scalability and low-cost nature made it an ideal choice for large data centers and Internet companies like Google and Amazon, gradually capturing a significant share of the global server market.

Additionally, FreeBSD\cite{quarterman19854}, released in 1993 as an open-source operating system based on Unix, performed excellently in the server market. FreeBSD inherited the core characteristics of Unix and further optimized network performance, file systems, and multitasking capabilities, becoming the preferred choice for high-performance network servers and embedded systems.

Entering the 21st century, the rapid development of virtualization technology and cloud computing fundamentally altered server operating systems' design and deployment methods. The maturity of hardware virtualization technology enabled servers to run multiple virtual machines, thus improving the utilization of hardware resources. Consequently, operating systems needed to support more efficient resource scheduling and isolation mechanisms. Operating systems like Linux and Windows Server\cite{WindowsServer} continuously optimized their kernels and virtualization support functions, meeting the needs of enterprises in data centers and cloud platforms. Particularly, Linux, the foundation for many cloud computing platforms, leveraged its lightweight, stable, and flexible characteristics to gain widespread application in public and private clouds.

\section{Recent Advances in OS Evolution}
\label{sec: Recent Advances}
With the continuous advancement of hardware technology, operating systems are undergoing unprecedented transformations. In recent years, there has been a notable enhancement in the computational power of CPU and GPU, which has propelled the development of high-performance computing and graphics processing technologies and provided a solid foundation for emerging fields such as AI and ML. Concurrently, Neural Processing Units (NPUs), specialized processors optimized for AI computations, are increasingly becoming standard configurations in many modern devices, further accelerating the advent of the intelligent computing era. However, with the widespread adoption of heterogeneous computing architectures, such as "big.LITTLE" designs, operating systems face new challenges, particularly task scheduling. Efficiently managing different types of processors to ensure tasks run on the most suitable hardware is an urgent issue that needs to be addressed.

Echoing this trend, the development of software technologies, such as Large Language Models (LLMs), also imposes new demands on operating systems. As complex systems capable of understanding and generating natural language, LLMs require substantial computational resources for training and operation. These models are running in data centers and beginning to migrate to edge devices, indicating that future operating systems must be more flexible to support a wide range of application scenarios. Furthermore, with the proliferation of IoT devices, distributed computing has become a trend, requiring operating systems to manage local resources and effectively coordinate remote resources across networks.

However, against the backdrop of rapid advancements in both software and hardware, existing operating systems have revealed specific inadequacies and difficulties. On the one hand, traditional operating systems' design philosophies and technology stacks struggle to keep pace with hardware technology, especially in terms of performance optimization and power consumption control. On the other hand, as user demands for security and privacy protection continue to rise, existing operating systems also face severe challenges. To adapt to future trends, operating systems must overcome current limitations and evolve into smarter, safer, and more efficient architectures to meet the ever-changing technological requirements. In this process, balancing performance, power consumption, security, and user experience becomes a critical topic in the evolution of operating systems.

\subsection{Operating Efficiency}

\subsubsection{Scheduling}

In contemporary operating systems, scheduling policies are pivotal in ensuring the efficient utilization of system resources and optimizing performance. However, as technology evolves and application demands diversify, traditional operating system scheduling strategies encounter many challenges and issues, particularly in process scheduling, heterogeneous resource allocation, load balancing, and energy efficiency optimization.

\textbf{1) Process scheduling}

Traditional heuristic scheduling methodologies, including First-Come, First-Served (FCFS), Shortest Job First (SJF), and Round-Robin (RR), have shown limitations in meeting the optimal scheduling demands within today's intricate operational landscapes. While these methods can be efficient under certain conditions, their inability to effectively manage tasks with diverse priorities and resource needs often leads to suboptimal resource allocation, diminished system performance, and prolonged response times. The advent of AI, especially advancements in ML, has prompted researchers to investigate applying these sophisticated tools to solve process scheduling problems, aiming to transcend the constraints posed by conventional approaches and bolster system performance.

Significant achievements have emerged from this area of inquiry. For example, the study "Comparative Analysis of Process Scheduling Algorithm using AI models" \cite{moni2022comparative} presents an innovative adaptive dynamic round-robin (ADRR) algorithm, which refines process scheduling by modulating time slice durations according to predicted CPU burst times. The findings reveal that the ADRR algorithm markedly diminishes average waiting times and substantially alleviates the problem of process starvation. Another investigation, "Improvement of lottery scheduling algorithm based on machine learning algorithm" \cite{yang2022improvement}, targets the refinement of the algorithm through the integration of ML to forecast process turnaround times and recalibrate weight distributions in lottery scheduling. This enhancement curtails average waiting periods and amplifies the operational system's efficiency, thus enriching the user experience.

Moreover, scheduling real-time tasks in cloud settings confronts additional layers of complexity. "A Deep Reinforcement Learning-Based Preemptive Approach for Cost-Aware Cloud Job Scheduling" \cite{cheng2023deep} suggests a preemptive strategy grounded in deep reinforcement learning to enhance the cost-benefit ratio of job scheduling in cloud platforms. By dynamically modifying the sequence of job executions, this technique ensures compliance with real-time stipulations while curtailing expenses. Similarly, the article "DRLBTSA: Deep reinforcement learning based task-scheduling algorithm in cloud computing" \cite{mangalampalli2024drlbtsa} unveils a task-scheduling algorithm termed DRLBTSA, which harnesses a deep Q-learning network model to refine the scheduling efficacy of tasks in cloud computing ecosystems. This methodology allows the system to ascertain more judiciously when and where tasks should be executed, thereby augmenting overall efficiency and responsiveness.

\textbf{2) Heterogeneous Resource Allocation}

The proliferation of multi-core processors and heterogeneous computing resources necessitates more intelligent oversight and coordination by operating systems. The architecture of heterogeneous cores, characterized by disparate performance traits, complicates scheduling efforts, as extant algorithms may falter in adapting to such diversity, resulting in imbalanced resource allocation and diminished scheduling proficiency. Nonetheless, the evolution of ML and allied advanced algorithms has spurred the exploration of novel solutions to these dilemmas.

"A reinforcement learning based job scheduling algorithm for heterogeneous computing environment" \cite{song2023reinforcement} outlines a dual-stage scheduling algorithm that utilizes bidirectional graph convolutional networks for initial task selection, succeeded by a heuristic technique that amalgamates optimistic cost tables and task duplication for processor assignment. This strategy facilitates superior utilization of varied resources within heterogeneous computing environments, elevating overall scheduling effectiveness. Concerning comprehensive resource management within operating systems, "SmartOS: towards automated learning and user-adaptive resource allocation in operating systems" \cite{goodarzy2021smartos} unveils a reinforcement learning-driven method that empowers operating systems to autonomously learn and dynamically regulate the distribution of CPU, memory, input/output (I/O), and network bandwidth according to user preferences. This innovation paves the way for creating more intelligent, user-centric operating systems.

In the realm of microservice management within cloud infrastructures, "$\mu$ConAdapter: Reinforcement Learning-based Fast Concurrency Adaptation for Microservices in Cloud" \cite{liu2023muconadapter} introduces the $\mu$ConAdapter framework, which exploits reinforcement learning to swiftly pinpoint and modify the ideal soft resource configuration for pivotal microservices, thereby reducing breaches of service level objectives (SLOs) and securing service excellence. Additionally, "Using machine learning techniques to analyze the performance of concurrent kernel execution on GPUs" \cite{carvalho2020using} illustrates the application of ML techniques, such as XGBoost, to dissect GPU benchmark suites, pinpointing essential resource requirement attributes that impact the performance of concurrent kernel execution, and optimizing the partitioning of GPU resources in heterogeneous setups.

Ultimately, "Operating systems for resource-adaptive intelligent software: Challenges and opportunities" \cite{liu2021operating} deliberates on achieving intelligent adaptation to fluctuations in heterogeneous resources via resource disaggregation, service-oriented resource provisioning, and learning-based resource scheduling and allocation. This addresses present-day challenges and opens up fresh vistas and prospects for developing future operating systems.

\textbf{3) Load Balancing}

Load balancing is a cornerstone technology for optimizing resource utilization and boosting system throughput in distributed systems and data center environments. Traditional scheduling strategies, however, frequently fall short in managing dynamically changing workloads. This can result in specific nodes being overloaded while others remain underutilized, negatively impacting overall system performance and stability. Recognizing these challenges, researchers have increasingly focused on integrating ML and other advanced algorithms into load balancing practices to foster more intelligent and efficient resource management.

In the paper "Machine Learning for Load Balancing in the Linux Kernel" \cite{chen2020machine}, the authors propose an ML-based resource-aware load balancer that mirrors the principles of the Completely Fair Scheduler (CFS). By leveraging extensive training datasets gathered from real-world operations, this method effectively manages compute-intensive workloads and mitigates resource contention, improving system response times and ensuring the equitable distribution of resources. Consequently, this approach enhances system stability and reliability and sets a precedent for applying ML to traditional scheduling mechanisms.

Another contribution to the field comes from the study "Batch Jobs Load Balancing Scheduling in Cloud Computing Using Distributional Reinforcement Learning" \cite{li2023batch}, which introduces a groundbreaking load balancing scheduling algorithm grounded in Distributional Reinforcement Learning. This innovative method learns the distribution of cumulative rewards through quantile regression, enabling it to adjust task allocations dynamically. As a result, it achieves superior cluster load balancing and boosts the efficiency of task completion, offering valuable insights and practical solutions for load balancing in expansive cloud computing environments.

Addressing the complexities of heterogeneous computing systems, the paper "A Machine Learning-Based Resource-Efficient Task Scheduler for Heterogeneous Computer Systems" \cite{hayat2023machine} presents a load balancing task scheduler enhanced with an ML-based device predictor. This scheduler employs ML to forecast the load conditions of various devices and assign tasks accordingly, effectively tackling the challenge of load imbalance in heterogeneous computing environments. Doing so not only elevates resource utilization but also significantly improves the system's overall performance, underscoring the potential of ML in optimizing resource management across diverse computing platforms.

\textbf{4) Energy Efficiency Optimization}

As High Performance Computing (HPC) becomes increasingly pervasive in areas such as scientific research, engineering simulation, and big data processing, the demand for high-density computational power continues to rise, and thermal dissipation efficiency and energy efficiency have become critical factors that cannot be ignored. Moreover, in energy-limited settings, such as mobile devices and IoT devices, extending device battery life is paramount in system design. Traditional scheduling methods, often lacking in consideration for energy consumption, not only lead to energy wastage but also hinder the optimal performance of devices. Therefore, developing low-energy consumption scheduling strategies to prolong device battery life has emerged as a significant challenge for modern operating system scheduling.

Researchers have started exploring innovative approaches that combine high-performance computing requirements with energy efficiency optimization to tackle this challenge. One such example is presented in the paper "Automatic Energy-Efficient Job Scheduling in HPC: A Novel SLURM Plugin Approach" \cite{aaen2023automatic}, which introduces an energy-efficient job scheduling method specifically designed for high-performance computing environments. This method involves the development of a new SLURM plugin that utilizes application-specific energy models to guide job scheduling decisions, effectively reducing energy consumption while preserving computational performance.

In the rapidly evolving domain of federated learning, the paper "Scheduling Algorithms for Federated Learning With Minimal Energy Consumption" \cite{pilla2023scheduling} delves into the optimization of energy consumption during training on heterogeneous devices by refining workload distribution. The authors propose an optimal solution strategy based on the multiple-choice minimum cost maximum knapsack packing problem and introduce four algorithms tailored for scenarios with monotonically increasing cost functions. These methods promote energy conservation and enhance resource utilization in distributed learning contexts.

Furthermore, in the context of Industry 4.0 and IoT environments, the paper "Heterogeneous Energy-Aware Load Balancing for Industry 4.0 and IoT Environments" \cite{ahmed2022heterogeneous} proposes an ML-based resource-aware processor selection method. This approach achieves effective load balancing in heterogeneous clusters and significantly reduces both execution time and energy consumption by assigning tasks to energy-efficient cores. These advancements underscore the importance of integrating energy efficiency into scheduling strategies across diverse computing platforms, reflecting a broader trend toward sustainable and optimized computing practices.

\subsubsection{Memory}

In the realm of memory management within modern operating systems, the growing complexity of workloads and the exponential increase in data volumes have rendered traditional static memory allocation strategies insufficient for achieving efficient resource utilization. To address these challenges, AI technologies, particularly ML, are increasingly integrated into memory management to elevate intelligence and efficiency.

Zhang et al. \cite{zhang2022software} introduced a Software-Defined Address Mapping (SDAM) mechanism that leverages ML to automatically detect program access patterns and optimize data placement within 3D stacked memory. This mechanism enhances bandwidth utilization and overall system performance and provides robust support for data-intensive applications, improving memory access efficiency.

Lagar-Cavilla et al. \cite{lagar2019software} proposed a software-defined remote memory scheme for warehouse-scale computing systems. This scheme proactively compresses infrequently accessed cold memory pages, effectively creating a remote memory layer. Doing so increases the available memory capacity and reduces energy consumption through diminished migration frequency between hot and cold data, thus optimizing memory usage and energy efficiency.

The Kleio project \cite{doudali2019kleio} represents a hybrid memory page scheduler that combines hierarchical storage management based on historical data with intelligent data placement decisions informed by deep neural networks. Kleio optimizes data layout for heterogeneous memory architectures, further enhancing the response speed and efficiency of the memory subsystem. This approach ensures that data is placed in the most appropriate memory tier, improving overall system performance.

In memory allocation, LLAMA \cite{maas2024combining} is an innovative memory allocator that uses ML to predict the lifecycle of objects, optimizing memory utilization rates and reducing fragmentation in large-page memory. By accurately predicting object lifecycles, LLAMA improves memory allocation efficiency and reduces the performance overhead associated with frequent allocation and deallocation operations, leading to more effective memory management.

Lastly, the Adaptive Huge-Page Subrelease strategy \cite{maas2021adaptive} focuses on optimizing the performance of non-migratory memory allocators in warehouse-scale computers. This strategy dynamically determines when to split large pages and return them to the operating system, enhancing overall performance. It introduces a new metric called "actual fragmentation" to measure the impact on large page coverage when applications rapidly release and reacquire memory, providing a more nuanced understanding of memory fragmentation and its effects on system performance.

These advancements in memory management highlight the potential of AI and ML in addressing the complexities of modern computing environments, leading to more efficient and intelligent memory utilization.

\subsubsection{I/O}

ML technology is emerging as a potent tool in operating system I/O optimization. Researchers have begun exploring utilizing ML to enhance the I/O subsystem within operating systems to accommodate modern, complex, and dynamic workloads.

The SmartOS project \cite{goodarzy2021smartos} introduces a reinforcement learning-based resource management approach, enabling the operating system to automatically learn and dynamically adjust the allocation of CPU, memory, I/O, and network bandwidth based on user priorities. This method enhances the system's adaptive capabilities and facilitates optimal resource allocation according to real-time demands.

The LinnOS study \cite{hao2020linnos} focuses on the unpredictable performance issues of flash memory storage, employing a lightweight neural network to infer the performance of SSDs. This has achieved predictable performance for storage applications and significantly reduced I/O latency, crucial for improving the response time and user experience of databases and cloud storage services.

In the application of ML, a study proposes a method to accelerate I/O \cite{serizawa2019accelerating} by overlapping data replication and reading, using local storage in high-performance computing clusters to alleviate the I/O bottleneck caused by large-scale training datasets. This approach is particularly suitable for ML frameworks such as Chainer, enhancing their reading bandwidth and the performance of data-parallel training.

Furthermore, addressing the remote access issue of GPU in AI applications, a study \cite{wang2024characterizingnetworkrequirementsgpu} proposes a GPU-centric method that determines the minimum network latency and bandwidth requirements for remote GPU invocation, ensuring that AI applications do not suffer in performance and may even be enhanced.

Lastly, the KML framework \cite{akgun2022using} is another attempt to integrate ML into the operating system's I/O subsystem, aiming to improve I/O performance and support adaptive configuration through ML. This framework provides the operating system with a flexible approach to handling diverse I/O demands, especially in scenarios requiring high throughput and low latency.

These studies demonstrate the potential of ML in optimizing operating system I/O. They address issues such as resource allocation, performance prediction, data handling, and network requirements from various perspectives and propel the development of operating systems toward greater intelligence.

\subsection{Security}

\subsubsection{ Threat Identification and Intervention}

Threat identification and intervention are indispensable pillars in safeguarding the security of information systems. They are crucial for the rapid detection and response to potential threats, thereby preserving data integrity, availability, and confidentiality. With the escalation in software complexity and the proliferation of diverse attack vectors, traditional security strategies struggle to cope with contemporary threats. Thus, harnessing advanced technologies, notably ML and automation, to augment threat identification and intervention capabilities has emerged as a central theme in ongoing research.

In software engineering, "Automated program repair: a step towards software automation" \cite{roychoudhury2019automated} delineates a methodology employing ML to automate the debugging process. This innovation targets reducing time spent on error correction, tackling the enduring issues faced in software development, and the complexities of collaboration within distributed teams. Automation facilitates the enhancement of both the quality and security of software products.

For Deep Learning (DL) models, "Toward actionable testing of deep learning models" \cite{xiong2022toward} outlines a comprehensive testing strategy aimed at identifying property violations or vulnerabilities. This approach guarantees these models' reliability, security, and robustness, laying a sturdy groundwork for their deployment in real-world applications. Rigorous testing serves to pinpoint and resolve inherent flaws, mitigating the risk of failures in critical operations.

In the context of the IoT, "A novel insider attack and machine learning based detection for the Internet of Things" \cite{chowdhury2021novel} identifies a new type of insider attack exploiting the RPL protocol's vulnerabilities, known as the sinkhole attack. It proposes a security framework utilizing ML to recognize unusual behavior patterns, thereby detecting and preempting such attacks and fortifying the security of IoT ecosystems.

Addressing the lifecycle management of IoT devices, "Reboot-oriented IoT: Life cycle management in trusted execution environment for disposable IoT devices" \cite{suzaki2020reboot} presents the RO-IoT framework. This framework employs autonomous updates of entire operating system images and leverages a Trusted Execution Environment (TEE) along with Public Key Infrastructure (PKI) certificates to manage the lifecycle of IoT devices, ensuring their security and dependability.

To combat software vulnerabilities, "VulRepair: A T5-based automated software vulnerability repair" \cite{fu2022vulrepair} introduces VulRepair, a method that leverages the T5 model and Byte Pair Encoding (BPE) for automatic vulnerability repairs. This technique enhances repair accuracy and feasibility, alleviating the burden on security professionals and accelerating the remediation process.

In malware detection for Android devices, "MSN-droid: The Android malware detector based on multi-class features and deep belief network" \cite{qin2019msndroid} integrates multiple feature layers, including native layer, permission, and system API features, with a Deep Belief Network (DBN). This integration results in precise malware detection, boosting the efficacy and precision of such detection mechanisms through a layered analytical approach.

For stealthy "Living-Off-The-Land" (LOL) attacks, "Living-off-the-land command detection using active learning" proposes an active learning framework called LOLAL. This framework selects uncertain and anomalous samples iteratively, incorporating analyst feedback. It demonstrates effective detection even with minimal labeled data, thus enabling swift identification and prevention of these covert attacks.

To detect mobile malware, "Dynamic detection of mobile malware using smartphone data and machine learning" \cite{panman2022dynamic} offers a method that applies dynamic hardware features and ML classifiers like random forests to identify Android-based mobile Trojans. Despite lacking privileged access, this method achieves commendable classification results, providing a viable solution for malware detection on mobile platforms.

Lastly, "Patching locking bugs statically with Crayons" \cite{cruz2023patching} describes a static automated program repair technology that integrates static analysis with ML to address misuse of sequential locking APIs in the Linux kernel. This technology markedly improves repair success rates by identifying potential locking errors via static analysis and suggesting fixes through ML, bolstering the resilience and security of kernel code.

\subsubsection{ Privacy protection}

Privacy protection is a critical pillar within contemporary information technology, committed to shielding personal information from unauthorized disclosure or misuse throughout data collection, storage, and processing. As the pace of digital transformation quickens, the significance of privacy-preserving technologies has grown, compelling researchers to delve into diverse methodologies to fortify privacy safeguards while concurrently enabling effective data utilization.

In the context of federated learning, the work "PPFL: Enhancing privacy in federated learning with confidential computing" \cite{mo2022ppfl} delineates an innovative ML framework that synergizes federated learning, differential privacy, TEEs, and multi-party computation techniques. This amalgamation is designed to mitigate privacy risks associated with centralized data aggregation by enabling model training on distributed datasets without directly exchanging original data. Consequently, this approach diminishes the likelihood of data breaches, bolstering privacy protections for users while maintaining the precision of predictive models.

Complementarily, the study "An efficient method for analyzing widget intent of Android system" \cite{qi2021efficient} presents a methodological advancement by integrating DL architectures such as MobilenetV3 and BiLSTM to analyze user intent through the extraction of both visual and textual data. This methodology enhances the accuracy of privacy-related detections and streamlines the training process, thus offering a more personalized and secure service to users without compromising their privacy.

Furthermore, "DarkNetz: towards model privacy at the edge using trusted execution environments" \cite{mo2020darknetz} explores a strategy that leverages TEEs alongside model partitioning on edge devices to minimize the potential attack surface of deep neural networks. By conducting sensitive computations locally, this technique mitigates the necessity for transmitting sensitive data over networks, thereby curtailing the risk of data exposure. The architecture of DarkNetz ensures heightened security for both models and data while preserving the performance integrity of the models in question.

\subsection{Structure Optimization}
Operating system architecture optimization aims to provide a more flexible and modular organization that can adapt to evolving technological demands and increasingly complex computational environments. Traditional operating system architectures are becoming increasingly limited in modern computing challenges, such as resource management in cloud environments, efficient construction of unikernels, and personalized user interaction requirements. Therefore, researchers are exploring new methods and technologies to enhance the flexibility and modularity of operating systems.

One notable advancement involves leveraging ML to augment user interaction within operating systems. Zhang et al. \cite{zhang2024enhanceduserinteractionoperating} have introduced an innovative framework that integrates LLMs with ML algorithms and interactive design principles to emulate authentic user behavior. This methodology facilitates the establishment of credible virtual A/B testing environments and propels the ongoing refinement of personalized services, ensuring that operating systems are more intuitive and responsive to user needs.

Another significant development pertains to the simplification of unikernel customization and application portability. Kuenzer et al. \cite{kuenzer2021unikraft} have pioneered the Unikraft platform, which adopts a fully modularized approach to operating system primitives and furnishes a suite of versatile, high-performance APIs. By doing so, Unikraft empowers developers to construct and deploy specialized unikernels with relative ease, bolstering the resulting systems' security and efficiency.

Moreover, advancements in kernel-bypass techniques have been instrumental in shaping the future of network stack design. Chen et al. \cite{chen2018survey} have conducted a comprehensive survey of user-space network stack configurations and evaluated various kernel-bypass methodologies. Their findings offer critical insights that guide the development of next-generation network stacks, enabling developers to craft high-performance applications with enhanced network capabilities.

The utilization of unikernels in the context of ML has also garnered attention. Leon et al. \cite{leon2020darkunikernelsmachinelearning} have explored strategies to optimize the unikernel construction process, capitalizing on the intrinsic security and performance attributes of unikernels to facilitate ML inference tasks. This line of inquiry highlights the untapped potential of unikernels in accelerating ML applications.

Addressing the intricacies of cloud resource management, Pemberton \cite{pemberton2021restless} has devised a technique to synthesize a cohesive "cloud system interface" from disparate cloud provider APIs. This synthesis encapsulates the core principles of interface design and delivers a standardized approach to managing resources across multiple cloud platforms, streamlining the orchestration of large-scale computing operations.

Zhang et al. \cite{zhang2021demikernel} have further contributed to the domain by introducing the Demikernel architecture, a specialized operating system design tailored for microsecond-scale datacenter systems and heterogeneous kernel-bypass devices. Demikernel's lightweight and adaptable nature supports seamless integration with existing systems, achieving sub-microsecond latencies and significantly boosting operational efficiency.

The fusion of unikernel optimizations with mainstream operating systems represents another frontier in OS innovation. Raza et al. \cite{raza2023unikernel} have proposed the Unikernel Linux (UKL) project, which seeks to amalgamate the performance and security benefits of unikernels with the extensibility and versatility of Linux. This initiative underscores a balanced approach to balancing specialized performance gains with broad compatibility and functionality.

Cadden et al. \cite{cadden2020seuss} have addressed the performance bottlenecks associated with serverless computing by introducing SEUSS, a system designed to expedite function deployment and caching through unikernel snapshots and page-level sharing mechanisms. SEUSS's design principles contribute to the efficiency and responsiveness of serverless architectures, aligning with the growing demand for scalable and agile cloud services.

Skiadopoulos et al. \cite{skiadopoulos2021dbos} have conceptualized DBOS, an operating system paradigm centered around a distributed transactional database management system (DBMS). This novel architecture serves as a robust foundation for scalable cluster operations. It integrates essential functionalities like scheduling, file management, and inter-process communication, catering to the needs of data-intensive applications.

These contributions redefine the boundaries of operating system design, fostering a landscape where systems are more adaptable, secure, and finely tuned to meet the specific demands of modern computing workloads.

\subsection{OS for AI}
Operating Systems for Artificial Intelligence (OS for AI) aim to integrate advanced AI technologies to optimize system performance, enhance user experience, and drive innovation in computing. With the evolution of AI technologies, particularly the advancements in LLMs, the design of operating systems has begun to incorporate more AI elements to achieve a more intelligent and personalized computing environment. These innovations improve the system's flexibility and efficiency and provide new perspectives for designing future computer systems.

A notable example is the AIOS-Agent ecosystem proposed by Ge et al. \cite{ge2023llmosagentsapps}, which envisions a future where LLMs form the bedrock of intelligent operating systems, facilitating the creation of versatile AI agent applications through natural language interfaces. This concept represents a paradigm shift in how we conceptualize and interact with computer systems, moving towards a more intuitive and dynamic user environment.

Building upon this vision, Mei et al. \cite{mei2024aiosllmagentoperating} delve deeper into the practical implementation of an AI-powered operating system named AIOS. By embedding LLMs directly into the OS, AIOS achieves enhanced capabilities in resource management, context switching, and security, which are critical for supporting the concurrent operation of multiple AI agents. This approach not only improves the computational efficiency and responsiveness of the system but also sets a new standard for the seamless integration of AI functionalities into everyday computing tasks.

Packer et al. \cite{packer2024memgptllmsoperatingsystems} address the challenge of expanding the contextual awareness of LLMs by introducing the MemGPT system, which leverages a sophisticated virtual context management mechanism. Drawing inspiration from the hierarchical memory management techniques used in conventional operating systems, MemGPT dynamically allocates and manages different types of memory to extend the effective context window of LLMs. This enhancement is crucial for enabling LLMs to perform more complex and context-sensitive tasks, thereby broadening the scope of applications for AI-driven systems.

Wu et al. \cite{wu2024oscopilotgeneralistcomputeragents} present the OS-Copilot framework, which is geared towards the creation of generalist digital assistants capable of autonomously interacting with the operating system and executing a wide range of tasks. The framework incorporates self-improvement algorithms that allow these digital agents to learn from and adapt to user interactions over time, leading to continuous performance enhancements. Such capabilities underscore the potential of AI to personalize and refine the user experience, making computing more accessible and efficient.

Xing et al. \cite{xing2023promptsapperllmempoweredsoftware} introduce Prompt Sapper, a novel infrastructure for developing AI-native software services that harnesses the power of LLMs. Prompt Sapper streamlines the development process, empowering developers—even those without specialized AI knowledge—to create sophisticated AI applications. This democratization of AI development tools is expected to spur innovation and accelerate the adoption of AI technologies across various industries.

Hè et al. \cite{hè2024perospersonalizedselfadaptingoperating} focus on personalization and privacy in the cloud-computing era with the introduction of PerOS, a personalized, self-adapting operating system. PerOS integrates LLM capabilities to offer a tailored user experience while ensuring robust data protection measures. The adaptive nature of PerOS allows it to evolve in response to individual user behaviors and preferences, thus delivering a more intuitive and secure computing environment.

Lastly, Hong et al. \cite{hong2023metagptmetaprogrammingmultiagent} propose MetaGPT, a meta-programming framework that leverages LLMs to optimize collaboration among multiple agents. By encoding operational protocols into prompt sequences, MetaGPT facilitates the efficient coordination of multi-agent systems, thereby enhancing their collective problem-solving capabilities. This framework exemplifies the potential of LLMs to transform the landscape of distributed computing and collaborative AI systems.

\section{New Era of OS}
\label{New Era of OS}
\subsection{Introduction}

In the rapidly evolving landscape of computing, the traditional paradigms of OS are being challenged by the emergence of new technologies and the changing demands of users. The advent of the IoT, cloud computing, and AI has not only expanded the scope of what an OS must manage but also introduced a host of new challenges and opportunities. Real-time processing, distributed computing, and enhanced security are becoming increasingly critical, necessitating a rethinking of how OSes are designed and implemented. The need for systems that can efficiently handle vast amounts of data, support complex and dynamic environments, and integrate seamlessly with emerging technologies is more pressing than ever. This section explores the new era of operating systems, focusing on the innovative approaches and architectures that are being developed to address these challenges. We examine how the integration of AI and ML is driving advancements in predictive analytics, adaptive resource management, and intelligent security. Furthermore, we discuss the concept of an ACOS, which proposes a modular, adaptable, and cross-platform compatible design, aiming to revolutionize the way OSes interact with hardware, software, and users. By abstracting system components into autonomous agents, ACOS seeks to achieve a flexible and scalable architecture that can adapt to various resource platforms, thereby enhancing system efficiency and user experience. This section synthesizes the current landscape and sets the stage for a forward-looking discussion on the future trajectories of operating systems, highlighting open issues and areas ripe for further research and innovation.

\subsubsection{ Challenges in Cross-Platform Adaptation}

Few current OSs can run on most devices across all categories, with corresponding distributions often requiring tailored optimization based on device characteristics. The vast disparity in resource platforms from high-performance servers to extremely resource-constrained embedded devices limits the capability of a single OS to cover all scenarios comprehensively. For most operating systems, supporting a new resource platform typically necessitates substantial modifications to core components, which not only increases development complexity but may also lead to a series of compatibility issues. Consequently, creating an OS that can seamlessly adapt to all computing devices remains a significant challenge, thereby restricting the practical emergence and widespread application of truly cross-platform operating systems.

The considerable differences between resource platforms necessitate that OSs consider specific hardware environments and application scenarios during the design phase. Otherwise, maintaining consistency and efficiency across different types of devices becomes difficult. This highlights that current OS designs still require optimization for specific hardware configurations to achieve optimal performance, reflecting the limitations of operating systems in terms of cross-platform versatility.

\subsubsection{ Poor Strategy Adaptability}

Although operating systems have been optimized for specific hardware environments and application scenarios during their design stages, there are inevitable discrepancies between actual hardware configurations, usage scenarios, and preset parameters. Most current operating systems manage resources and execute tasks using heuristic-based global static strategies, which prove inadequate in addressing these discrepancies, leading to suboptimal optimization outcomes.

While such static management strategies may be optimal under preset conditions, they lack flexibility in dynamic real-world environments. When hardware configurations or workloads change, static strategies cannot promptly adjust to new circumstances, thus affecting overall performance. Therefore, the effectiveness of optimization solutions in existing operating systems is limited when facing diverse and continuously changing operational conditions.

\subsubsection{ Difficulties in Cross-Domain Collaboration}

In today's IoT and cloud computing environment, ineffective collaboration mechanisms among devices and systems from different domains lead to low data exchange and resource-sharing efficiency. As smart devices and cloud services become more prevalent, cross-device and cross-platform data sharing and collaboration are becoming increasingly important. However, software for device collaboration often needs to be developed specifically for each domain, significantly increasing the complexity and cost of system integration.

In most cases, interoperability between devices is constrained by their respective operating systems and communication protocols, making resource sharing challenging. For instance, in a smart home scenario, devices from different manufacturers might use various communication standards such as Wi-Fi, Bluetooth, Zigbee, etc., and their operating systems could differ. This means that dedicated adapters or middleware must be written for each pair of devices to achieve seamless collaboration between these devices, which is time-consuming and labor-intensive.

Furthermore, the lack of standardized data formats and communication protocols can result in data loss or misinterpretation due to format mismatches, even when physical connections are feasible. This issue is particularly evident in Industrial IoT (IIoT), where inefficient information sharing among sensors, robots, and other automated equipment within a factory can lead to reduced production efficiency and increased operational costs.

Another issue is redundant computation. When devices cannot effectively share computational resources, each must possess full computational power and storage capacity, wasting resources and potentially causing performance bottlenecks. For example, in edge computing scenarios, if edge devices cannot borrow computational capabilities from one another, individual devices may fail to respond to requests on time under high load conditions due to insufficient processing power.

\subsubsection{ Challenges in User Interaction}

With the advancement of the Internet of Everything, the variety of devices that user software needs to support has become increasingly diverse, highlighting the growing disparity between the development of interaction devices with significant differences and the human resources required for software interaction development. The enrichment of user software by Large Language Models (LLMs) and AI Agents further exacerbates this disparity, constraining the quality and user experience of interactions within systems. Specifically, user interaction faces the following challenges:

First of all, diverse devices require different interaction. Modern smart devices vary widely, each posing unique challenges for user interface design. Smartwatches, with tiny screens, require simple and efficient interfaces using swipe gestures, button clicks, and voice commands. Smartphones, with moderate screens, need interfaces that balance information density and usability for touch operations. Desktop computers, with larger screens, can handle complex interfaces with multiple windows and detailed settings but must avoid overwhelming users. Smart TVs, used at a distance, need large fonts and icons for clear visibility and easy remote control operation. Developing distinct interfaces for each device type is labor-intensive and complicates maintaining consistency, increasing development and maintenance costs. User habits and expectations also vary across devices, making it challenging to create a universal user interface design.

Furthermore, modern software requires personalization and dynamic functionality. With the advancement of technology, especially LLMs and AI Agents, systems can automatically combine tools and learn user behaviors, generating new applications and expanding functionalities. These new, personalized, and auto-generated tools require corresponding user interfaces to support their functions. However, existing user interface generation technologies often fail to adapt quickly to the needs of these new tools, leading to poor user experiences.

Another issue is the limited flexibility and intuition. Traditional interaction methods are insufficiently flexible and intuitive in many scenarios, affecting user operational experience. Using text or voice for simple inputs like switching states or selecting items can introduce unnecessary delays and complexity. Text alone struggles to intuitively express certain states, such as date selection or color picking. Additionally, the lack of animation and other user feedback mechanisms makes the results and responses generated by systems less intuitive and appealing.

\subsection{Next Generation OS Paradigm}

\subsubsection{Introduction}

With the rapid advancement of information technology, modern computing devices exhibit a trend toward diversification and heterogeneity. From high-performance servers and PCs to mobile devices and IoT terminals, these devices display significant differences in hardware configurations and vary widely in application scenarios and user requirements. For instance, high-performance servers typically need to handle large-scale datasets and high-concurrency requests, imposing stringent computational power and storage capacity demands. In contrast, mobile devices prioritize power management and user experience, requiring smooth operation and extended battery life within limited resources. Embedded devices must maintain stable operation in extreme environments while possessing real-time processing capabilities. This diversity and heterogeneity present unprecedented challenges for the design of OS.

The varying resource platforms and requirements have led to substantial differences in implementing core components across different device operating systems. These differences make it difficult to directly reuse modules, forcing developers to undertake independent development efforts for each OS environment. For example, a file system module that performs well on a server may not be directly transplantable to a mobile device due to the latter's stricter requirements for memory usage and power consumption. Similarly, a task scheduler designed for embedded devices may not meet the high-concurrency demands of a high-performance server. In such cases, developers must invest considerable time and effort into redesigning and implementing these modules to accommodate different hardware platforms and application scenarios.

Designing a highly modular, scalable, and maintainable operating system has become crucial to address these challenges. In this context, we propose the concept of ACOS. The central idea behind ACOS is to abstract all operating system components into independent agents. The OS provides the foundation and environment for these agents to meet user needs better, while the collaborative outcomes of the agents contribute to the healthy and efficient operation of the OS. This concept, encapsulated in our "Agent for OS, OS for Agent" philosophy, aims to enhance the flexibility and adaptability of the operating system.

By adopting the ACOS approach, we envision an operating system that dynamically adapts to various hardware configurations and application scenarios. Each agent can be optimized for specific tasks and environments, allowing for efficient resource utilization and improved performance. Furthermore, the modular nature of ACOS facilitates more manageable maintenance and updates, enabling the OS to evolve alongside technological advancements and user needs. This innovative design paradigm holds the potential to revolutionize the field of operating system development, paving the way for more versatile and efficient computing solutions.

\subsubsection{Agent Abstrcation}

In the face of increasingly diverse and complex computing environments, the significant differences in hardware configurations, application scenarios, and user requirements across various devices pose substantial challenges to the reusability of core components in OSs. For instance, the resource management needs of servers and mobile devices differ markedly; servers emphasize high-concurrency processing capabilities, whereas mobile devices focus on power consumption control and user experience. These disparities increase developers' workload and limit operating systems' flexibility and adaptability.

To address these challenges, ACOS proposes the concept of "Anything Anywhere All As Agent," which abstracts all operating system components—kernel modules, drivers, or user-space applications—into agents. Through this conceptual abstraction, ACOS can:

\begin{itemize}[leftmargin=1em, itemindent=0em, labelindent=0em]
\item \textbf{Enhance System Modularity:} 
By abstracting functional modules into agents, ACOS achieves a highly modular system architecture. This simplifies replacing and combining components, promotes loose coupling between different components, and enhances the system's scalability and maintainability.

\item \textbf{Improve System Adaptability:} 
The loosely coupled design of AI agents allows the OS to flexibly adjust its composition based on the characteristics of different resource platforms, ensuring efficient operation across multiple devices. Additionally, the intelligent attributes of AI agents enable them to better adapt to environmental changes, enhancing the overall adaptability of the system.

\item \textbf{Promote Cross-Platform Compatibility:} 
The agent abstraction facilitates the easy migration of similar functionalities across different platforms, significantly reducing cross-platform application porting costs and improving software generality and portability.

\item \textbf{Simplify Function Expansion and Optimization:} 
In ACOS, adding or optimizing specific functions requires only introducing or replacing relevant AI agents without extensive modifications to the entire system. Moreover, the ability to dynamically load and unload AI agents further enhances system flexibility, enabling rapid responses to new requirements and technological changes.

\end{itemize}

Currently, AI agents are typically defined as intelligent entities capable of perceiving environmental changes, making decisions, and taking action to achieve specific goals. Key attributes of AI agents include autonomy, reactivity, sociality, adaptability, reliability, and security 
Considering the broad range of device adaptations and dynamic functional needs of ACOS, AI agents in ACOS must also satisfy the following additional characteristics:

\begin{figure*}[!t]
\centering
\includegraphics[draft=false, width=\textwidth]{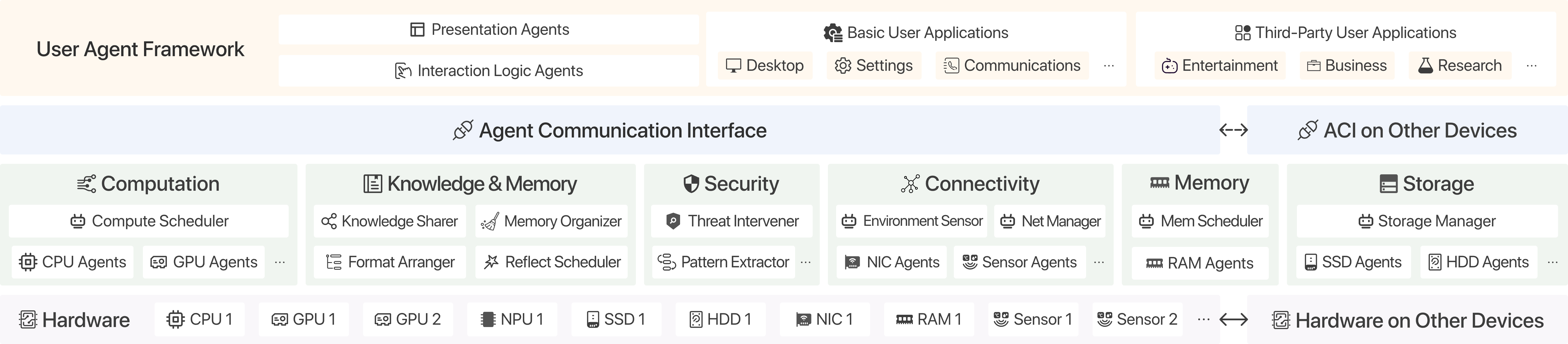}
\caption{\textbf{ACOS Architecture.} The figure depicts the modular architecture of the ACOS. It showcases a layered design where the User Agent Framework interacts with various applications and services, while the Agent Communication Interface (ACI) facilitates inter-agent communication. Specialized agents manage computation, knowledge, security, connectivity, memory, and storage, integrating seamlessly with underlying hardware components. This structure supports efficient resource management and intelligent system operations.}
\label{ACOS Architecture}
\end{figure*}

\begin{itemize}[leftmargin=1em, itemindent=0em, labelindent=0em]
     
\item \textbf{Composability:}
Composability refers to the ability of AI agents to dynamically combine and decompose based on task requirements and environmental changes, thereby achieving more efficient and flexible task execution and resource allocation. This characteristic enables multiple AI agents to flexibly integrate into a higher-level agent to accomplish newly emerging tasks. The resulting composite agent possesses enhanced processing capabilities and a broader perception range and forms a unified entity by collaborating with its constituent sub-agents, thereby achieving new collective objectives.

Conversely, when faced with simple or specific tasks, a complex agent can be decomposed into multiple independent sub-agents, each focusing on handling particular tasks or functions. This approach improves resource utilization efficiency and enhances the system's reliability and fault tolerance. By dynamically adjusting the composition of AI agents, the system can optimally allocate resources and adapt to varying operational conditions, ensuring robust and efficient performance.

\item \textbf{Instrumentality:}
Instrumentality refers to the capability of an individual agent to serve as the most basic unit for completing independent tasks and to function as a tool or resource for another agent, providing specific functionalities or services. This characteristic enables AI agents to form complementary relationships, which leverage each other's capabilities to accomplish their tasks more effectively. Instrumentality is not limited to hardware agents but can also be applied to software agents, such as those providing computational, storage, or communication services.

Instrumentality enhances the system's resource utilization and functional extensibility by fostering inter-agent cooperation. It ensures that each agent can fully utilize its strengths to support and service other agents, optimizing the system's overall performance and efficiency.

\item \textbf{Collaborativeness:}
Collaborativeness refers to the ability of multiple AI agents to effectively communicate and coordinate with one another to accomplish tasks collectively. This characteristic enables AI agents to form teams, divide labor, and collaborate to achieve common goals. Collaborativeness requires agents to possess robust communication capabilities and the ability to understand and adapt to other AI agents.

Through collaboration, AI agents can share information, coordinate actions, and resolve conflicts, thereby enhancing the overall efficiency and effectiveness of the system. Collaborativeness is a critical feature for the efficient operation of multi-agent systems, as it allows the system to leverage each agent's capabilities and resources fully. This synergy ensures the system can optimize its performance and achieve its objectives more effectively.

\item \textbf{Scalability:}
Scalability refers to the ability of a multi-agent system to expand its capabilities and performance by incorporating new agents or functional modules. This characteristic enables the system to continually upgrade and adapt to changing requirements and technological advancements, maintaining long-term viability and competitiveness. Scalability requires the system to have a well-designed modular architecture and requires flexible interfaces and protocols to facilitate the seamless integration and collaborative operation of new agents.
\end{itemize}

Through scalability, the system can continuously adapt to new application scenarios and technological developments, ensuring it meets future needs. This ongoing adaptability is crucial for sustaining the system's relevance and effectiveness in dynamic and evolving environments.

\subsubsection{ACOS Architecture}

ACOS adopts a flat architectural design, treating all system modules, software, and hardware drivers abstracted as agents with equal structural status. This design enables ACOS to manage and utilize various computing resources more effectively while enhancing the system's stability and reliability.

Precisely, the architecture of ACOS consists of several key components:

\begin{itemize}[leftmargin=1em, itemindent=0em, labelindent=0em]

\item \textbf{Agent Communication Interface (ACI):} As an intermediary interface, the ACI connects different types of agents, facilitating efficient communication between them. This ensures that all agents can collaborate on an equal structural footing. The ACI achieves higher communication efficiency by eliminating the overhead associated with hierarchical designs.

\item \textbf{Agents:} In ACOS, agents consist of APP/Shell, Kernel, and Hardware Agents.

App/Shell Agents manage user interactions and provide user-facing functionalities; Kernel Agents handle the management of underlying system resources; Hardware Agents control hardware devices and provide low-level hardware management.

\item \textbf{Hardware:} The direct hardware infrastructure on which ACOS operates.

\end{itemize}

ACOS's system architecture inherently supports distributed collaboration. Through various physical links at the hardware level, the communication interface of ACOS can be extended to all devices within the same connected environment. This significantly expands the scope of agent collaboration, enhancing the possibilities and potential for task coordination among agents.

\subsection{Agent for OS}

\begin{figure*}[!t]
\centering
\includegraphics[draft=false, width=400px]{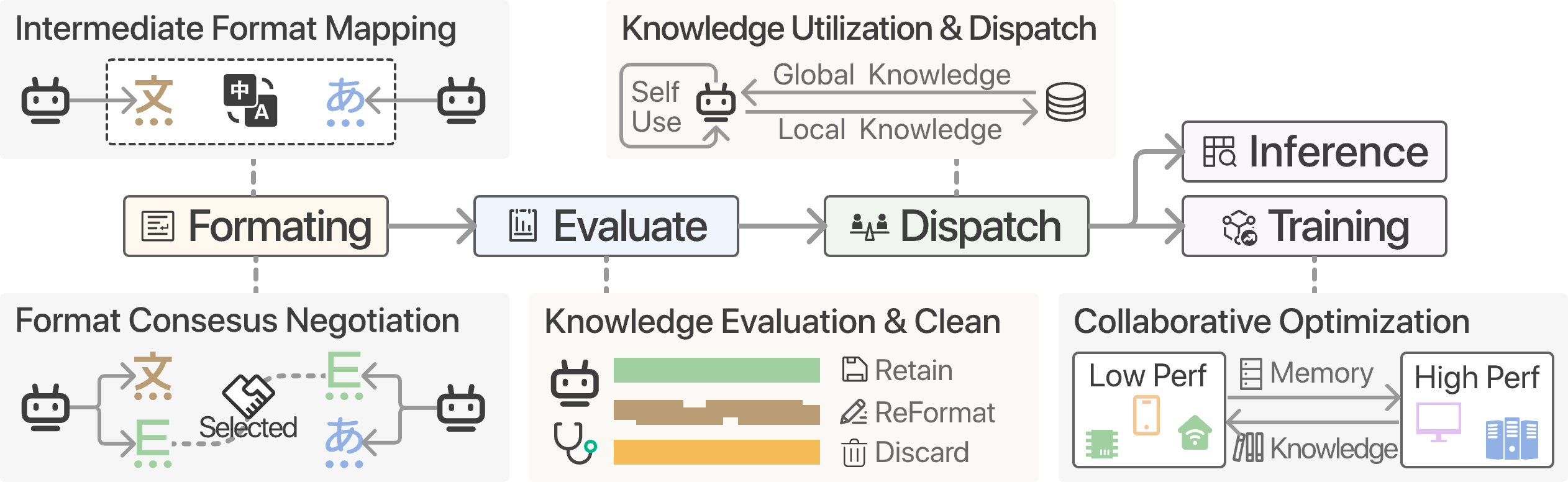}
\caption{\textbf{ACOS Knowledge \& Memory Management.} This figure illustrates the knowledge and memory management framework within the ACOS. It demonstrates the process of unifying intermediate formats, evaluating and dispatching knowledge, and optimizing collaborative performance. The diagram shows how agents negotiate for format consensus, retain, reformat, or discard information, and utilize local and global knowledge to achieve high-performance optimization.}
\label{ACOS Knowledge Memory Management}
\end{figure*}
\subsubsection{Knowledge \& Memory Manager Agent}

Due to their varying roles, different agents often exhibit significant differences in the forms of knowledge and memory they manage. For example, a network monitoring agent might record detailed network traffic data and patterns. In contrast, a task scheduling agent might focus on the execution characteristics and resource usage of tasks on the device. Suppose knowledge and memory are stored haphazardly without proper management. In that case, they will become difficult for other agents to utilize, leading to information silos among agents and affecting the efficiency and capability of knowledge utilization within the system.

One of the critical features of agents in ACOS is that they are knowledge-based. A lack of high-quality knowledge can result in poor agent performance, while a lack of high-quality memory can hinder knowledge improvement, impacting the user experience and efficiency of system usage. Therefore, it is essential to manage and schedule the knowledge and memory generated by system operations, such as logs, to enhance the capabilities of agents within the system continuously. Specifically, the following aspects require management:

\textbf{1) Knowledge and Memory Formats}

Different agents manage their knowledge and memory in ways that reflect their primary tasks and capabilities. This variation makes it difficult for one agent to understand and utilize the knowledge and memory of another agent to optimize its knowledge. To break down the isolation of knowledge and memory among agents, when introducing new agents, it is necessary to negotiate and collect the formats of their persistent knowledge and memory, ensuring they adopt formats that align with the system-wide standards. Additionally, mappings should be established between these formats and the system-wide standard formats to enable the system to comprehensively understand the state of knowledge and memory, facilitating subsequent integrated utilization.

\textbf{2) Storage Forms}

The differences in agent tasks lead to variations in how they store knowledge and memory. For instance, some agents require high-frequency, fast-recovery log storage, while others manage knowledge and memory that consume significant storage space. ACOS should meet these differentiated needs by managing the storage of agent knowledge and memory, thereby improving the efficiency of production and utilization and facilitating management and application.

\textbf{3) Log Processing}

During the operation of agents, a large volume of logs is generated, many of which lack the value necessary for knowledge utilization. To enhance the efficiency of log utilization, the Knowledge Manager will perform data cleaning, format conversion, removal of redundant information, and annotation and evaluation of the logs. Specifically, data cleaning involves removing invalid or erroneous log entries; format conversion ensures that all logs conform to a uniform standard; the removal of redundant information prevents the unnecessary occupation of storage space; and annotation and evaluation involve adding tags and weights to necessary log entries to facilitate subsequent analysis and utilization. Through these steps, the Knowledge Manager can significantly improve the quality and usability of log data, providing robust support for the intelligent management and optimization of the system.

\textbf{4) Update Scheduling}

Agents have varying knowledge requirements. Some agents, such as those performing static data analysis, may not require frequent knowledge updates due to the nature of their tasks. Conversely, agents responsible for dynamic environment sensing or customer service may need more frequent updates to adapt quickly to environmental changes or user needs. The devices on which agents reside also influence knowledge updates; for example, edge computing devices, constrained by computational power, often cannot perform frequent updates. Therefore, ACOS should be able to schedule knowledge updates and optimizations for agents based on their characteristics and the knowledge and memory generation within the system.

\textbf{5) Knowledge Selection}

When optimizing knowledge, an agent's internal knowledge and memory may have limitations, leading to suboptimal results in multi-agent collaborations. Therefore, knowledge selection and infusion are necessary to help agents focus on the characteristics of their environment, ensuring that the behavior goals of agents within the system are aligned. ACOS should select appropriate portions of knowledge from the knowledge bases of other managed agents and provide them as supplementary knowledge to enhance agents' comprehensive capabilities and development.

\subsubsection{Compute Scheduler Agent}

\begin{figure*}[!t]
\centering
\includegraphics[draft=false, width=400px]{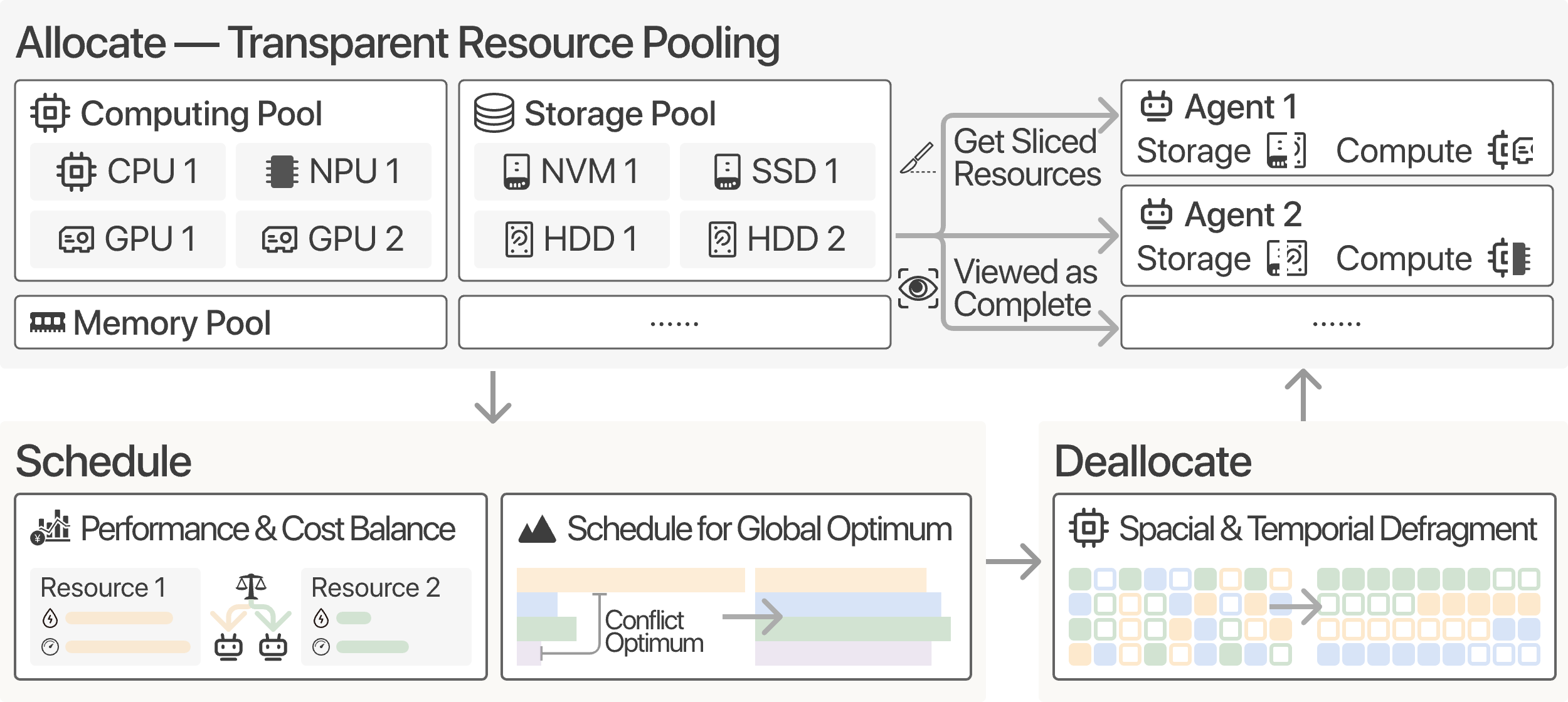}
\caption{\textbf{ACOS Resource Managing Manicheism.} Figure illustrating the concept of Agent Centric Resource Scheduling in an operating system environment, demonstrating how resources from computing pools (CPU, NPU, GPU and so on) and storage pools (NVM, SSD, HDD and so on) are allocated transparently to individual agents based on their requirements, followed by scheduling strategies aimed at optimizing performance and cost balance, spatial-temporal defragmentation, and achieving global optimum resource utilization.}
\label{ACOS Resource Managing Manicheism}
\end{figure*}

In modern computing environments, the operating system, the core program responsible for managing computer hardware and software resources, is crucial in delivering high-performance computing services. The operating system's efficiency and flexibility are essential for optimizing resource utilization. With the widespread adoption of multi-core processors and the advancement of heterogeneous computing architectures, the available resource types within a system have expanded significantly. Traditional process scheduling mechanisms based on fixed rules are now facing unprecedented challenges.

The Compute Scheduler Agent is designed to efficiently integrate and manage internal computing resources within the system to address evolving workload patterns and the demand for multi-objective performance optimization. This goes beyond traditional CPU time-slice allocation and extends to the comprehensive scheduling of various computing units, including GPU, NPU, and DPU. By adopting a holistic and in-depth resource integration strategy, the Compute Scheduler Agent can more flexibly adapt to different computational demands, ensuring that the hardware potential is maximized across various application scenarios.

The Compute Scheduler Agent must collaborate closely with computational hardware agents such as CPU, GPU, NPU, and DPU Agents to create an efficient and coordinated working environment. This involves not only identifying and aggregating all available computing resources but also having the capability to allocate these resources effectively to different tasks. Particularly in handling large-scale parallel computing tasks, the Compute Scheduler Agent can employ optimization algorithms to facilitate effective collaboration among multiple computing units, thereby enhancing the overall computational efficiency of the system. Additionally, by abstracting heterogeneous computing resources into a unified compute pool, the Compute Scheduler Agent can dynamically adjust resource allocation strategies based on the specific characteristics of each task, ensuring that every task receives the most suitable computational support for its execution conditions. This approach improves the overall computational efficiency of the system.

The Compute Scheduler Agent focuses on optimization in the following areas:

\textbf{1) Resource Utilization} 

Enhancing the system's overall efficiency by minimizing idle times for resources. This means that when tasks are waiting to be executed, the scheduler should be capable of rapidly identifying and allocating appropriate resources to these tasks, thereby avoiding performance wastage due to resource idleness. Furthermore, effective resource allocation strategies can promote efficient collaboration among tasks for multi-core and heterogeneous computing environments, improving the system's overall performance.

\textbf{2) Dynamic Adaptability} 

The scheduler must possess high flexibility and self-adjustment capabilities to adapt to application requirements and system load changes. This requires scheduling policies to quickly respond to external environmental changes, such as the addition of new tasks or changes in the priority of existing tasks, without compromising system stability. Dynamic adaptability ensures the system can consistently provide efficient service under various operating conditions, particularly in highly uncertain and dynamic application scenarios.

\textbf{3) Energy Efficiency Optimization} 

While ensuring performance requirements are met, measures are taken to minimize unnecessary energy consumption. This involves strategically scheduling the use of computational resources during task execution, such as dynamically adjusting processor frequency and voltage to save power or optimizing the physical layout of tasks to reduce communication overhead. Effective energy management can extend operational time and enhance user experience for battery-powered devices. In large-scale data centers, it can help reduce operational costs and carbon emissions.

\textbf{4) User Intent Fulfillment} 

Deeply understanding and predicting user habits and preferences to align scheduling decisions more closely with actual user needs. For instance, in personal computing environments, the scheduler can quickly respond to user actions, such as launching applications, and prioritize these tasks to ensure they receive sufficient computational resources. In enterprise applications, resources should be flexibly allocated based on the characteristics of business processes to support the efficient operation of critical business tasks. By doing so, the system can improve response times and processing capabilities and significantly enhance user satisfaction and work efficiency.

To provide superior scheduling choices compared to traditional rule-based heuristics, the Compute Scheduler Agent must acquire knowledge about the system and processes, enabling device-specific and environment-adapted scheduling decisions. This is reflected explicitly in the following aspects:

\begin{itemize}[leftmargin=1em, itemindent=0em, labelindent=0em]

\item  \textbf{Computational Unit Configuration:} 
Understanding the current device's computational architecture, including the number, type, and instruction sets of cores. For instance, for processors with multiple cores of different performance levels (such as "big.LITTLE" designs), the scheduler must be able to distinguish and fully utilize the characteristics of these cores to enhance parallel processing capabilities and system response speed.
\item  \textbf{Task Characteristic Analysis:}
Collecting and evaluating each process's requirements and current states, such as computational intensity, I/O operation frequency, and estimated execution time. The runtime status of processes and their associated tasks helps the scheduler make more timely and accurate decisions. For example, for tasks nearing completion, the scheduler can appropriately extend their time slices to release computational resources sooner; tasks performing critical computations should ensure they receive the necessary support to prevent starvation and other issues.

\item  \textbf{User Behavior Patterns:} 
Gaining a deep understanding of user usage habits, including the types of frequently run applications and peak workload periods. By continuously observing and learning user behavior, the scheduler can predict future resource demands and prepare accordingly. For example, suppose it is observed that users typically open email clients to check messages in the morning. In that case, the scheduler can pre-activate relevant services before this period to reduce startup time and enhance user experience.

\end{itemize}

The extensive coverage of device types in ACOS necessitates that the system operates effectively across platforms with significant performance variations. Consequently, the process scheduling methods must be meticulously designed to meet the high demands for real-time performance. For devices ranging from low to high performance, the tools available to the Compute Scheduler Agent include, but are not limited to:

\begin{itemize}[leftmargin=1em, itemindent=0em, labelindent=0em]
    
\item\textbf{Rule-Based Heuristic Scheduling:}
Optimizing scheduling parameters through device virtualization or leveraging the computational power within the environment to enhance performance.

\item\textbf{Multi-Scheduler Algorithm Coordination:}
Dynamically switching between different scheduling methods based on changes in load scenarios to meet dynamic requirements.

\item\textbf{AI Scheduling with ML/DL:}
Utilize lightweight ML and DL models to intelligently generate scheduling decisions and continuously update these decisions using knowledge within the system.
\end{itemize}

\subsubsection{Memory Manager Agent}
With the increasing complexity of computing architectures and the diversification of application requirements, traditional memory management mechanisms have gradually revealed several limitations. For example, fixed memory allocation strategies struggle to adapt to dynamically changing workloads, often leading to memory fragmentation and resource wastage. The lack of intelligent analysis of memory usage patterns results in sluggish system responses to sudden memory demands. Additionally, traditional memory management mechanisms often overlook the specific needs of different application scenarios, such as the distinct memory requirements of real-time and batch-processing tasks. These issues impact the system's overall performance and can lead to decreased system stability and degraded user experience.

To address these challenges, the Memory Manager Agent is an essential component in the ACOS architecture, aiming to improve memory resource utilization and overall system performance through intelligent decision-making. Specifically, the optimization goals of the Memory Manager Agent focus on the following areas:

\textbf{1) Memory Utilization Efficiency} 

Ensuring that memory resources are fully utilized and minimizing additional memory overhead caused by fragmentation and lifecycle inconsistencies. This means that when there is a memory request, the Manager can quickly and accurately allocate appropriately sized and positioned memory blocks to the requester. Simultaneously, it should promptly reclaim memory when it is no longer needed, enhancing the continuity and availability of memory space.

\textbf{2) Dynamic Adaptability} 

Responding to changes in system memory demands, such as application startups, shutdowns, or shifts in memory usage patterns. This requires the Manager to have self-adjustment capabilities, enabling it to dynamically modify memory management strategies based on the current system state and predicted future trends. For instance, the detection of increased system memory pressure can trigger memory compression or migration operations to alleviate memory congestion.

\textbf{3) System Stability} 

Optimizing memory usage while ensuring system stability and reliability are not compromised. This includes preventing system crashes or significant performance degradation due to improper memory management. Safe memory management strategies can effectively mitigate these issues, such as setting memory usage caps and reserving emergency memory regions.

\textbf{4) User Demand Responsiveness} 

Understanding and adapting to changes in user memory usage demands, especially in multi-user or multi-task environments. For example, graphic-intensive applications or big data processing tasks can be prioritized for more memory allocation, while lightweight tasks can be allocated less memory to achieve balanced resource distribution.

To achieve these goals, the Memory Manager Agent must possess the following characteristics:

\begin{itemize}[leftmargin=1em, itemindent=0em, labelindent=0em]

\item\textbf{Compute Resource Awareness:} Understanding the current device's memory architecture, including both physical and virtual memory's capacity, speed, and latency.

\item\textbf{Memory Demand Analysis and Negotiation:} Collecting and evaluating each process's actual memory requirements and expected behaviors, such as memory consumption rates and peak memory usage. By negotiating with agents regarding memory usage and occupancy duration, memory utilization efficiency can be improved.

\item\textbf{System State Monitoring:} Continuously monitoring the system's memory usage, including critical metrics such as total memory occupancy and free memory levels. It should execute memory compression and migration operations when necessary to maintain optimal system performance.

\end{itemize}

\subsubsection{Storage Manager Agent}

In today's complex computing ecosystem, storage management, serving as a critical bridge between applications and underlying hardware, has become increasingly significant. With technological advancements, the variety and quantity of storage media in modern systems have grown substantially, ranging from high-speed non-volatile memory (NVM) to SSDs, traditional HDDs, and even tape storage. Each medium possesses unique performance characteristics and cost-effectiveness. This diversity brings flexibility but poses new challenges in efficiently managing and utilizing these resources.

Storage Manager aims to construct a storage system that meets high-performance demands while ensuring high reliability. This is achieved through intelligent data placement strategies, efficient storage optimization mechanisms, and comprehensive data protection measures. To accomplish this goal, the Storage Manager Agent must transcend traditional single-tier storage solutions and adopt a more flexible and adaptable hierarchical storage architecture. This architecture allows data to be automatically placed on the most appropriate storage tier based on access patterns, importance, and other factors, thereby achieving an optimal balance between speed, capacity, and cost. Additionally, the architecture must ensure the smoothness and efficiency of data migration processes to minimize the system overhead associated with data movement.

Building on this foundation, the Storage Manager Agent must also possess high transparency, presenting a unified and continuous storage space view to upper-layer applications and other system components regardless of the actual storage tier. This transparency simplifies application design and enhances the overall flexibility and maintainability of the system. To achieve this, the Storage Manager Agent must effectively manage the abstraction layer of storage resources, ensuring that upper-layer applications do not need to concern themselves with the specific implementation details of the underlying storage.

To ensure data security and integrity, the Storage Manager Agent incorporates disaster recovery mechanisms that automatically execute multi-level backup strategies based on the importance of different data. These mechanisms include local redundant backups and may involve cross-geographical data replication to meet the "3 copies, 2 types of media, at least one offsite" (3-2-1) principle, thereby enhancing data availability and durability.

Since data access patterns can change over time, the Storage Manager Agent adopts dynamic data migration strategies. It regularly analyzes data's access frequency and importance, moving active data to faster storage tiers and infrequently accessed data to lower-cost media. This adaptive storage optimization process is ongoing and automatically adjusts according to changes in the system's internal and external environments, ensuring the system remains optimal.

Finally, the Storage Manager Agent strongly emphasizes data security in response to growing demands for data privacy protection. It implements strict data isolation measures and selective encryption techniques to ensure the security of sensitive information throughout its lifecycle. Whether data is stored at rest or transmitted, it is effectively protected against unauthorized access and leakage.

\subsubsection{Network Manager Agent}

Network management is paramount as a critical link connecting various computing nodes in network-intensive and distributed computing environments. With the rapid advancement of network technology and the increasing complexity of application scenarios, traditional network management strategies have become inadequate in meeting the modern computing system's demands for high bandwidth, low latency, and dynamic adaptability. Against this backdrop, the design philosophy of the Network Manager Agent focuses on establishing an intelligent and efficient network resource management system. The goal is to optimize data transmission paths, enhance network resource utilization, and improve the system's dynamic adaptability, thereby comprehensively elevating network performance.

The Network Manager Agent aims to address the challenge of effectively managing network resources in highly dynamic network environments, ensuring the efficiency and reliability of data transmission. It must possess robust network awareness to achieve this, enabling real-time monitoring of network conditions, including key performance indicators such as bandwidth, latency, and packet loss rates across various network interfaces. This deep network insight forms the foundation for the Network Manager Agent's intelligent decision-making, allowing it to flexibly adjust data transmission strategies based on current network conditions, thus ensuring optimal network service quality under any circumstances.

When handling multiple potential network links, the Network Manager Agent employs advanced traffic management and scheduling algorithms to intelligently select the most suitable transmission paths among multiple network interfaces. This process involves more than simply choosing the fastest or most stable link; it also considers overall network load balancing to avoid overloading a single link while considering the rate and latency requirements of data transmission to ensure efficient and smooth data flow through the network. Additionally, the Network Manager must be able to recognize and learn network patterns by analyzing historical data and current network activities, predicting potential network congestion or other anomalies, and taking preemptive measures to maintain stable network operations.

In response to the varying network resource requirements of different applications and services, the Network Manager Agent also undertakes the task of fine-grained traffic control. This means it must dynamically adjust network resource allocation based on factors such as applications' importance and real-time requirements, ensuring that critical applications receive priority data transmission channels while other applications are served according to the availability of remaining resources. Such a mechanism enhances the efficiency of network resource utilization and increases the system's flexibility and responsiveness, ensuring that even under suboptimal network conditions, critical tasks can still be completed successfully.

\subsubsection{Security Agent}

In modern computing environments, the continuous evolution of network attack techniques has rendered traditional security measures insufficient to meet the increasingly complex demands of network security. To address this, the Security Agent focuses on constructing an intelligent security defense system capable of proactively identifying potential threats and possessing high adaptability to counter evolving attack methods. By profoundly integrating ML and advanced data analytics, the Security Agent can learn the normal behavior patterns of various Agents within the system and effectively detect deviations from these norms, thereby assessing the presence of potential security threats.

To achieve this objective, the Security Agent must first be equipped with detailed behavioral knowledge, encompassing a wide range of known security threat characteristics and patterns. Through continuous monitoring of the activities of various Agents within the system, the Security Agent can continually collect new data and compare it with existing behavioral pattern knowledge. The Security Agent can promptly respond and implement appropriate protective measures when detecting anomalous behavior—whether a new variant of a known threat or an unknown novel attack. Furthermore, to enhance threat detection accuracy, the Security Agent can automatically extract useful features from vast amounts of historical data, continuously refining its detection models to identify unknown threats effectively.

Beyond strengthening its own threat detection capabilities, the Security Agent emphasizes collaborative work with other system components. In network environments, the information available from a single node is often limited. By sharing threat intelligence with other nodes, the overall security level of the entire network can be significantly enhanced. Therefore, the Security Agent is designed with an efficient threat intelligence exchange mechanism that allows rapid sharing of the latest threat information between devices of varying performance levels. Devices with weaker computational capabilities can obtain the latest security updates from higher-performance devices without bearing the heavy burden of data processing tasks. This collaborative mechanism improves the network's response speed to newly emerging threats. It promotes the widespread dissemination of cybersecurity knowledge across the network, forming a more tightly integrated security defense network.

\subsubsection{Environment Sensing Agent}
In developing intelligent environment sensing systems designed for broad device coverage, the Environment Sensing Agent focuses on integrating and coordinating Sensor Agents distributed across various geographical locations to capture and understand physical environmental characteristics precisely. The approach is rooted in the deep integration of edge computing and IoT devices within the ACOS environment. By leveraging these devices' inherent environmental sensing capabilities, the goal is to provide detailed and real-time data about the physical world for upper-layer applications and services.

The core of Environment Sensing Agent is constructing a perception network that can dynamically adapt to environmental changes. This network must identify and integrate data streams from different types of sensors and flexibly adjust the frequency, precision, and scope of data collection based on specific environmental requirements and conditions. To achieve this, Environment Sensing Agent must possess robust data processing and analysis capabilities to ensure that valuable information is extracted from massive sensor data. This involves understanding and parsing the information provided by each Sensor Agent, including but not limited to their location, sensing modality (such as temperature, humidity, light intensity), data format, and sampling rate.

Moreover, the design of Environment Sensing Agent must consider the effective management and maintenance of a large and heterogeneous group of Sensor Agents. Given the low-power characteristics of IoT devices, the collaboration with the Sensor Agents running on them must consider factors such as energy consumption and employ reasonable task scheduling and synchronization mechanisms. This allows Sensor Agents to autonomously or semi-autonomously adjust their operating modes based on environmental changes and the needs of upper-layer applications. This design allows Environment Sensing Agent to function as an intelligent sensing node, operating independently and collaborating with other environmental sensing nodes to form a multi-layered, multifunctional environmental sensing network.

To achieve comprehensive environmental state perception, Environment Sensing Agent must also possess advanced data analysis and pattern recognition capabilities. This goes beyond simple data aggregation to recognizing patterns and trends within the data and even predicting future environmental changes. Environment Sensing Agent can learn environmental behavior patterns from historical data, enhancing its predictive capabilities regarding environmental changes. This provides more accurate and timely data support for decision-making, resource management, and other upper-layer applications.

\subsection{OS for Agent}

\subsubsection{User Agent Interaction}

\begin{figure*}[!t]
\centering
\includegraphics[width=400px, draft=false, ]{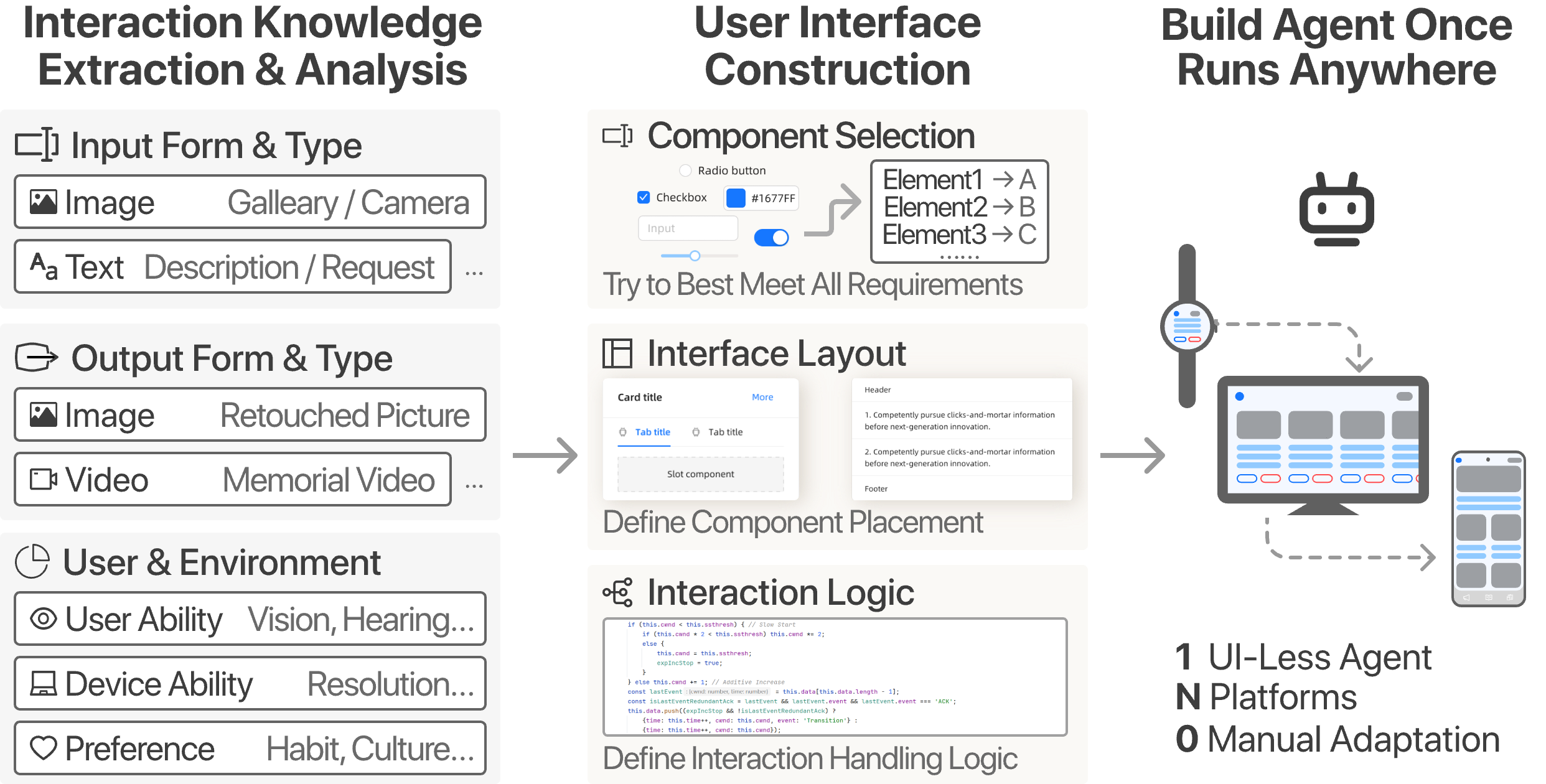}
\caption{\textbf{ACOS Agent UI/UX.} This figure presents an overview of the User Interface/User Experience (UI/UX) design considerations for ACOS Agents within the context of an ACOS. It highlights key aspects such as input/output forms and types, user interface construction including component selection, layout definition, interaction logic, and the principle of building once but running anywhere anytime. The diagram emphasizes the importance of tailoring the UI/UX to meet diverse requirements by considering factors like user abilities, device capabilities, preferences, habits, culture, etc., thereby enhancing usability and accessibility across different environments and scenarios.}
\label{ACOS Agent UI/UX}
\end{figure*}

To overcome the limitations of existing agent interaction methods and provide a more intuitive and user-demand-oriented interaction experience, ACOS proposes a brand-new interaction design concept. This design concept not only emphasizes the transition from CLI to GUI but also focuses on fully considering the characteristics of agents (such as input and output types) and user needs (such as the acceptance of size, color, and interaction methods by different age groups) with minimal user intervention to produce the most suitable GUI. To achieve this, ACOS has constructed a multi-layered design framework at its Shell layer, ensuring that every step, from information collection and analysis to the final construction of the user interface, is closely linked and highly automated.

\textbf{1) Knowledge Collection and Organization}

At the knowledge collection level, ACOS should thoroughly consider the requirements of various aspects of agent interaction and obtain and organize necessary information, focusing specifically on the following:

Input/Output Forms of Agents: Understanding the types of inputs and outputs supported by agents, such as text, images, sound, etc. This helps determine the most suitable user interface components for the current interaction scenario. For example, if an agent primarily handles image recognition tasks, the interface may need to integrate camera controls and an image preview area.

Characteristics and Habits of User Interaction: Recording typical interaction methods used by users, such as touchscreen gestures, mouse clicks, etc. Additionally, personalization settings such as font size, weight, accessibility features, and text direction must be considered. During system operation, user behavior data and A/B testing methods can be used to learn user preferences, which can be applied to improve the interaction interface. This information can help optimize the layout and control design of the user interface, enhancing user operational efficiency.

Specific Characteristics of Devices: Screen resolution, touch capabilities, physical button configurations, and other characteristics differ among desktop computers, laptops, tablets, and smartphones. Based on these characteristics, recommendations can be made for the most suitable user interface designs.

\textbf{2) User Interface Construction}

At the user interface construction level, appropriate interaction components are selected for interactive elements based on knowledge about agents, users, and devices, and layouts along with business logic are generated to form usable interaction interfaces.

This multi-layered design allows ACOS to handle complex interaction scenarios flexibly, providing users with smooth, intuitive, and personalized interaction experiences, whether for simple information queries or complex task executions. This design concept not only improves the efficiency of user-agent interactions but also greatly enhances the adaptability and accessibility of user interfaces, opening up new possibilities for future intelligent interaction technologies.

\subsubsection{Agent Security and Permission Management}

\begin{figure*}[!t]
\centering
\includegraphics[draft=false, width=\textwidth]{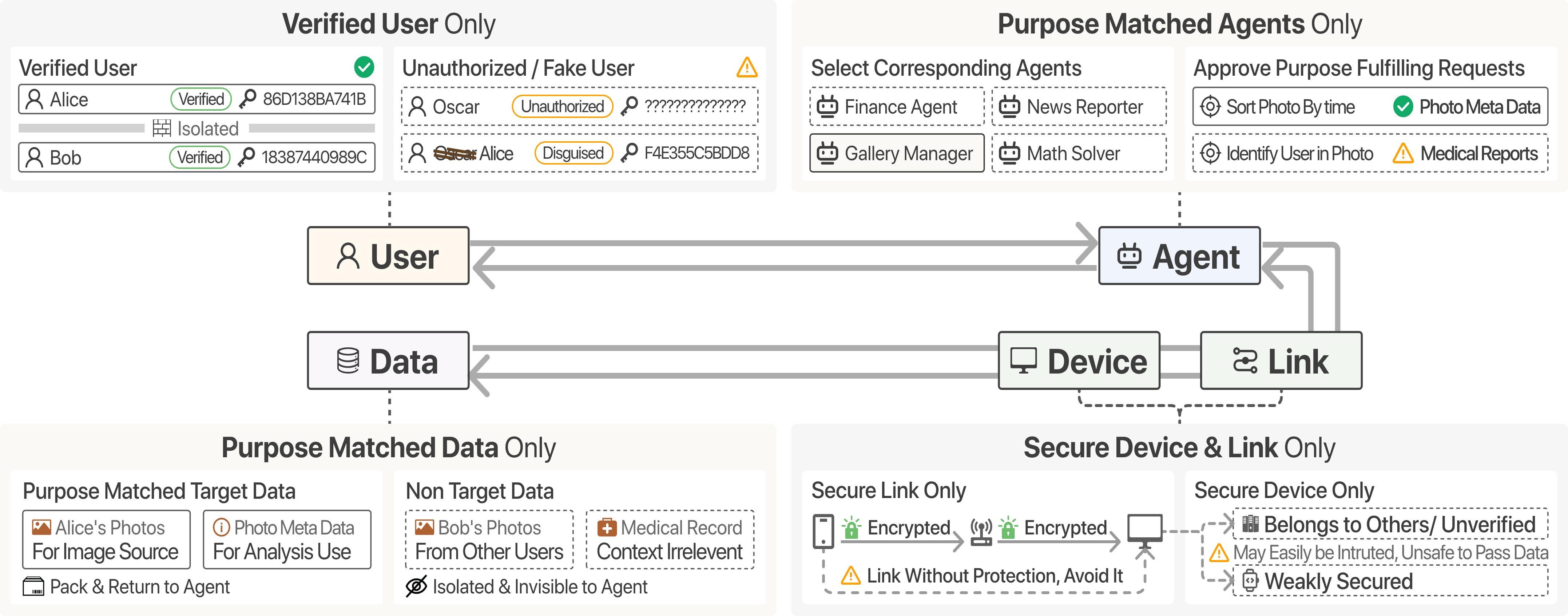}
\caption{\textbf{ACOS Secure Datapath.} This diagram illustrates the secure data path within an ACOS, highlighting its ability to ensure user data security across all stages of the process. It begins with verifying users before allowing them access to specific agents tailored to their needs or purposes. These agents then communicate securely over encrypted links to devices that have been verified as safe and appropriate for handling sensitive information. Finally, purpose-matched target data is delivered back to authorized users via this protected pathway, ensuring confidentiality throughout the entire transaction.This diagram illustrates the secure data path within an ACOS, highlighting its ability to ensure user data security across all stages of the process. It begins with verifying users before allowing them access to specific agents tailored to their needs or purposes. These agents then communicate securely over encrypted links to devices that have been verified as safe and appropriate for handling sensitive information. Finally, purpose-matched target data is delivered back to authorized users via this protected pathway, ensuring confidentiality throughout the entire transaction.}
\label{ACOS Secure Datapath}
\end{figure*}

In ACOS, data flows through four aspects: from the user to the agent, then through a link to the device, to meet users' dynamic and diverse needs. Since ACOS supports various device types and facilitates native distributed agent collaboration, it faces a significant challenge: securely and efficiently transmitting data between aspects with differing ownership and security capabilities. To tackle this issue, ACOS emphasizes four core security and permission management characteristics. The goal is to create a data flow environment that is both flexible and secure. Specifically, each stage of the data flow should:

\textbf{1) Credibly Certified}

To prevent sensitive data from being illegally intercepted or tampered with, ACOS ensures that all stages of data flow undergo strict authentication processes. This means that a user, an agent, a link, or a device must all go through a series of authentication steps before participating in data exchange to verify their identity and legitimacy. By doing so, the system can prevent unauthorized entities from intervening, ensuring that data flows only through predefined, secure paths and safeguarding privacy.

\textbf{2) Permission Granted}

Unauthorized access to unpermitted data poses a threat to user privacy. Therefore, ACOS adopts a fine-grained permission management strategy to ensure that participants can only access and manipulate data relevant to their tasks. This permission management not only assigns permissions based on participants' roles but can also dynamically adjust according to task changes. Through this method, ACOS can ensure the regular operation of system functions while minimizing unnecessary data exposure, thus protecting user privacy from infringement.

\textbf{3) Meet Security Capability Standards}

Due to various constraints, the security capabilities of each stage in data flow often differ. For example, IoT devices may lack the computational power to run complex encryption algorithms, whereas computers often have security mechanisms such as Trusted Platform Modules (TPMs). If the stages through which data flows are not appropriately chosen, weak links in security may be introduced, posing potential risks of data leakage and destruction. Therefore, the system needs to evaluate the security capabilities of each device and the sensitivity of each task/data to decide which data and tasks can be scheduled to which devices. For susceptible data, the system prioritizes aspects with more robust security capabilities for processing, while for general data, more relaxed standards can be adopted to achieve the best balance between security and efficiency.

\textbf{4) Ensure Ownership Security}

In the distributed collaboration environment of ACOS, each stage may belong to different owners, and clearly defining ownership boundaries at each stage is a crucial means of maintaining data security. The system needs to identify and manage devices, agents, and links belonging to different owners, ensuring that data flow adheres to strict ownership rules. For instance, personal data should not leave the scope of personal control without permission, and enterprise-sensitive data should only be processed in secure environments within the enterprise, thereby effectively preventing data leakage risks due to unclear ownership.

\subsubsection{Agent Task Scheduling}

\textbf{\indent 1) Agent Perception Mechanism }

To mitigate the overhead associated with continuous Agent polling, ACOS facilitates Agents to subscribe to events pertinent to their operations. It is imperative that system events are detectable by the respective Agents, referred to as triggering Agents, to ensure timely action. ACOS employs an event-driven architecture coupled with temporal management to accomplish this. Whenever predefined conditions are satisfied—such as the receipt of an external command or the attainment of a scheduled time—the pertinent Agent is activated to undertake the requisite task. Post-task completion, the Agent transitions into a dormant state, ready for subsequent activations. This mechanism ensures Agents remain responsive to both internal and external stimuli, preserving the system's dynamism and adaptability.

\textbf{2) Agent Context Switching and Continuity}

When transitioning between operational states, ACOS meticulously maintains the runtime context of Agents, encompassing ongoing data processing, intermediate results, and configuration settings. This provision guarantees seamless continuation of tasks, even when Agents migrate across devices or undergo temporary hibernation, thus upholding service continuity and stability. By doing so, ACOS optimizes resource utilization, contributing to the system's operational efficiency.

\textbf{3) Cross-Device Scheduling Strategy}

To foster effective multi-Agent collaboration, ACOS incorporates a sophisticated cross-device scheduling framework. This framework dynamically relocates Agents to optimal devices based on environmental conditions and task specifications, thereby enhancing task fulfillment and user satisfaction. Specifically, the system evaluates the capabilities and attributes of available devices to determine the most suitable platform for executing a given task. For instance, computationally intensive tasks are directed to devices boasting superior processing power, whereas latency-sensitive tasks are delegated to edge devices with minimal delay.

\subsubsection{Agent Collaboration}
    
\textbf{\indent 1) Agent Topology Design}

The topology design within ACOS balances flexibility and efficiency, allowing Agents to form dynamic mesh networks at the physical layer, which ensures robust connectivity. Logically, these networks can be structured in star configurations centered around key nodes, facilitating streamlined control and management. This design principle ensures the system's adaptability to diverse application scenarios, whether they demand decentralized decision-making processes or centralized resource allocation.

Given the dynamic nature of collaboration networks, where Agents may join or depart unpredictably, ACOS features an adaptive mechanism for selecting central nodes. This selection process is automated and context-aware, ensuring the system's resilience against rapid network changes and Agent failures, thereby sustaining operational efficiency and service quality.

\begin{figure}[!t]
\centering
\includegraphics[draft=false, width=\columnwidth]{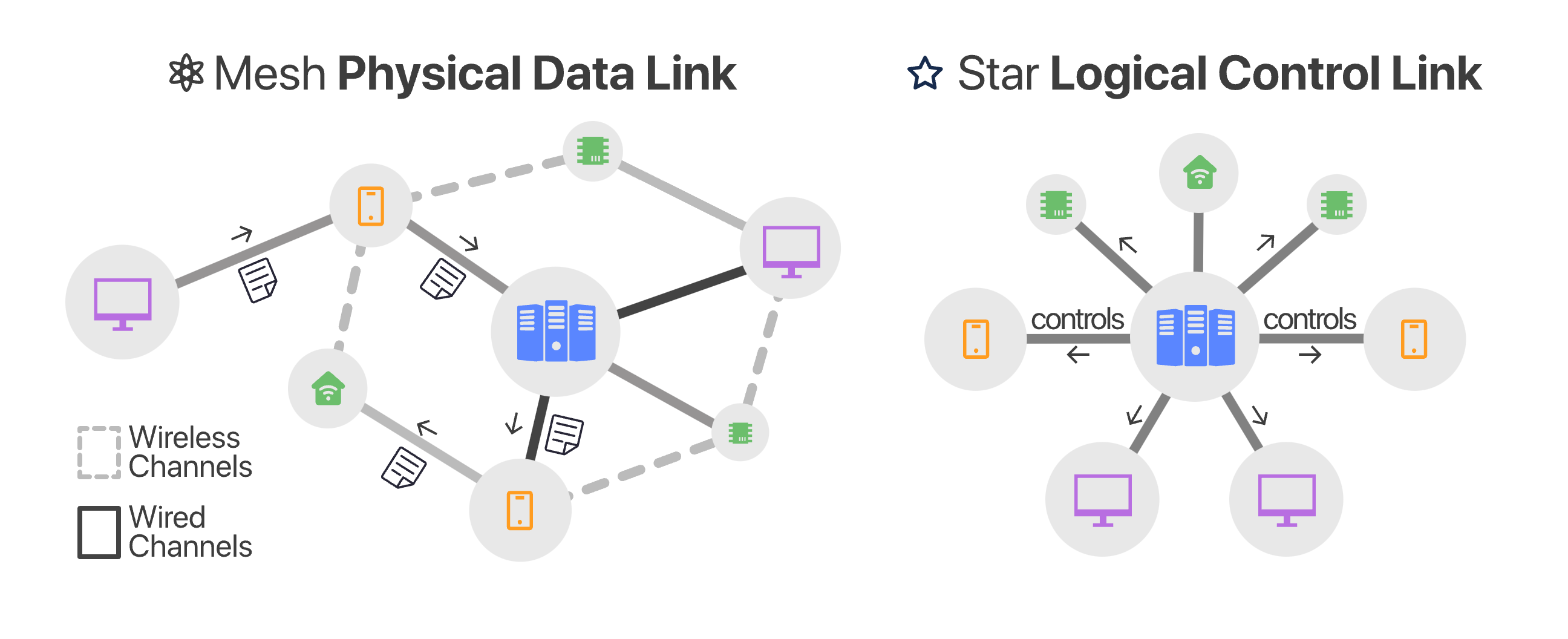}
\caption{\textbf{Agent Topology Structure.} The figure illustrates the ACOS topology, divided into two parts: Mesh Physical Data Link and Star Logical Control Link. The left side depicts the physical data link, where dashed lines represent wireless channels while solid ones indicate wired channels connecting various nodes in a mesh network configuration. On the right side is the logical control link, featuring a central node controlling multiple terminal devices' data transmission processes through star topology connections. Together, these components form the overall network architecture of ACOS.}
\label{Agent Topology Structure}
\end{figure}
\textbf{2) Collaboration Initiation}

Upon a new device's integration into the ACOS environment, the resident Agent promptly initiates a discovery protocol, notifying the surroundings of its presence via broadcasting or alternative communication channels. This process extends beyond mere physical proximity to include logical connections, such as shared data streams or aligned application objectives. The newcomer must seamlessly integrate into the established collaboration ecosystem, capable of exchanging information with existing Agents and presenting its capabilities and service interfaces to establish mutual trust rapidly. This mechanism supports continuous system expansion and optimization, reinforcing the system's openness and scalability.

For Agents encountering each other for the first time, ACOS mandates adherence to a standardized communication protocol. This ensures interoperability through uniform message formats and interaction protocols, irrespective of the Agents' origins or functionalities. During the initial encounter, the newcomer initiates a self-introduction, disclosing essential details about its capabilities. Established Agents subsequently evaluate this information to determine the newcomer's role within the collaboration network, aligning with their own operational needs. The overarching aim of this mechanism is to bolster the system's and Agents' adaptability to distributed and dynamic environments, thereby enhancing overall performance and user experience.

\section{Conclusion and Outlook}
\label{sec: Conclusion and Outlook}
In conclusion, the evolution of OS reflects the dynamic interplay between technological advancements and user needs. This review has traced the development from monolithic designs to modern microkernel architectures, highlighting significant milestones such as the rise of graphical user interfaces, mobile operating systems, and cloud-based platforms. These advancements have enhanced system functionality and usability, paving the way for new paradigms like the IoT and edge computing.

In recent years, the integration of AI and ML into OS has driven innovations in predictive analytics, adaptive resource management, and intelligent security. While these technologies offer substantial benefits, they also introduce challenges related to data privacy, transparency, and ethical considerations.

The ACOS represents a transformative approach to OS design, emphasizing modularity, adaptability, and cross-platform compatibility. By abstracting system components into autonomous agents, ACOS achieves a flexible and scalable architecture that can adapt to various resource platforms. This design simplifies maintenance and updates, enhances system efficiency, and improves user experiences across different devices.

ACOS addresses security and data management through fine-grained permission management, ensuring that data access is restricted to relevant tasks and minimizing unnecessary exposure. The system evaluates device security capabilities and task sensitivities to optimize data flow paths, balancing security and efficiency. Ownership security is maintained to prevent data leakage, ensuring data remains within appropriate control scopes.

Regarding the user interface, ACOS adapts to device-specific characteristics, providing smooth, intuitive, and personalized interactions. The multi-layered design enhances adaptability and accessibility, opening new possibilities for intelligent interaction technologies.

Future challenges for ACOS and similar systems include scaling agent communication networks, developing robust security frameworks, and addressing computational constraints in edge devices. Research should focus on these areas and integrate emerging technologies.

In summary, the future of operating systems will likely be characterized by greater intelligence, flexibility, and user-centric design. ACOS exemplifies this vision, offering a promising framework for the next generation of operating systems. By continuing to innovate and address evolving needs, the field can drive significant advancements, paving the way for a more connected, efficient, and secure computing environment.

\section{References Section}
\label{sec: Ref}

\bibliographystyle{IEEEtran}
\bibliography{ACOS}
\vfill

\newpage
\section{Acknowledgment}

The IoT picture in figure~\ref{Development of Embedded Operating Systems} was cited from \href{https://www.flickr.com/photos/wilgengebroed/8249565455/}{Wilgengebroed}, licensed under a \href{https://creativecommons.org/licenses/by/2.0/}{Creative Common Attribution 2.0 Generic License}

The first GPU picture in figure~\ref{Development of Desktop Operating Systems}, was cited from \href{https://commons.wikimedia.org/wiki/File:KL_TI_TMS34020.jpg}{Konstantin Lanzet}, licensed under a  \href{https://creativecommons.org/licenses/by/3.0/}{Creative Common Attribution 3.0 Generic License}

The first GUI picture in figure~\ref{Development of Desktop Operating Systems}, was cited from \href{https://commons.wikimedia.org/wiki/File:X-Window-System.png}{Liberal Classic}, licensed under a X11 License

The video game picture in figure~\ref{Development of Desktop Operating Systems} was cited from \href{https://www.flickr.com/photos/paulsynnott/2610499050/}{gwaar}, licensed under a \href{https://creativecommons.org/licenses/by/2.0/}{Creative Common Attribution 2.0 Generic License}

The RUST picture in figure~\ref{Development of Desktop Operating Systems}, was cited from \href{https://github.com/rust-lang/rust-artwork/blob/master/logo/rust-logo-blk.svg}{™/®Rust Foundation}, licensed under a \href{https://creativecommons.org/licenses/by/4.0/}{Creative Common Attribution 4.0 Generic License}

The i386 picture in figure~\ref{Development of Desktop Operating Systems}, was cited from \href{https://commons.wikimedia.org/wiki/File:80386SL_processor_from_1990.jpg}{Andre Celere}, licensed under a \href{https://creativecommons.org/licenses/by/4.0/}{Creative Common Attribution 4.0 Generic License}

The enterprise LAN in figure~\ref{Timeline of Key Developments in Server Operating Systems and Related Technologies} was cited from \href{https://commons.wikimedia.org/wiki/File:Wyższa_Szkoła_Filologiczna_computer_room.jpg}{BogdanWSF}, licensed under a  \href{https://creativecommons.org/licenses/by/3.0/}{Creative Common Attribution 3.0 Generic License}

\end{document}